\documentstyle[aps,eqsecnum,prd,epsf]{revtex}

\def\sqr#1#2{{\vcenter{\hrule height.#2pt
      \hbox{\vrule width.#2pt height#1pt \kern#1pt
          \vrule width.#2pt}
      \hrule height.#2pt}}}

\def\edth{{\rlap{$\partial$}\raise0.3em\hbox{$-$}}}
\def\cosec{{\rm cosec}}

\begin{document}

\vspace*{-10mm}
\epsfxsize=2cm
\leftline{\epsfbox{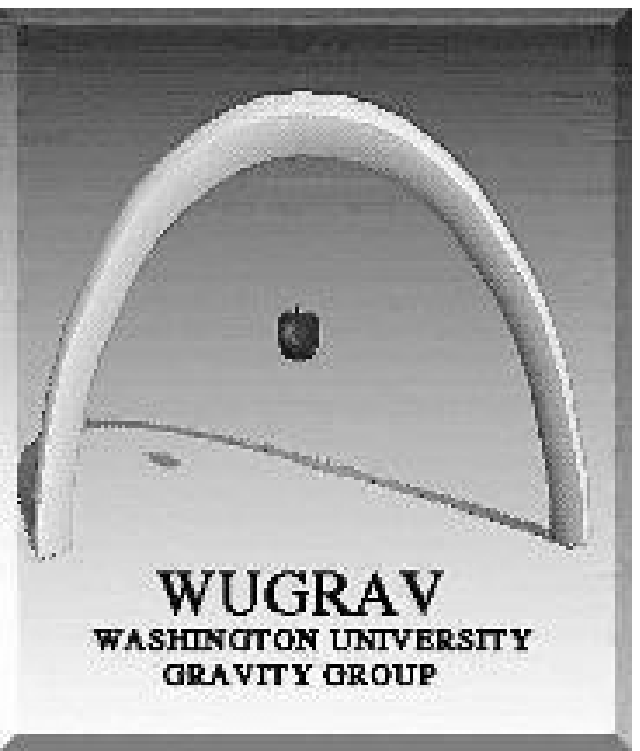}}
\vspace{5mm}

{\baselineskip-4pt
\font\yitp=cmmib10 scaled\magstep2
\font\elevenmib=cmmib10 scaled\magstep1  \skewchar\elevenmib='177
\leftline{\baselineskip20pt
\vbox to0pt
   { \hbox{{\yitp Washington \hspace{1.5mm} University 
\hspace{1.5mm} in \hspace{1.5mm} St}.\hspace{1.5mm}{\yitp Louis} }
     {\large\sl\hbox{{WUGRAV}} }\vss}}

\bigskip
\bigskip
\bigskip

\begin{center}

{\Large \bf Perturbative Approach to an orbital evolution 
around a Supermassive black hole}

\bigskip

{\large Yasushi Mino
\footnote{Electronic address: mino@wugrav.wustl.edu}}

\medskip

{\em Department of Physics, Washington University, Campus Box 1105, \\ 
One Brookings Dr., St.Louis, MO 63130-4899, USA}\\

\medskip

\today

\bigskip

{\bf abstract}

\end{center}

{\small 
A charge-free, point particle of infinitesimal mass 
orbiting a Kerr black hole 
is known to move along a geodesic. 
When the particle has a finite mass or charge, 
it emits radiation which carries away orbital energy and angular momentum, 
and the orbit deviates from a geodesic. 

In this paper we assume that the deviation is small 
and show that the half-advanced minus half-retarded field 
surprisingly provides the correct radiation reaction force, 
in a time-averaged sense, 
and determines the orbit of the particle.}

\bigskip


\section{Introduction}

Binary systems of solar-mass compact objects and supermassive black holes 
at galactic nuclei are expected to be 
important sources of gravitational waves. 
In order to detect these gravitational waves, 
a project to construct a space-based detector, LISA, is underway\cite{LISA}. 
The detection of such gravitational waves 
will reveal fundamental information about gravitational theory, 
and will provide insights into conditions at the centers of distant galaxies. 
It may be possible to read out 
the detailed geometrical structure of a supermassive black hole 
encoded in the incoming gravitational waves, 
with which we can test gravitational theory 
in the strong gravitational region, 
and our understanding of black holes. 
Not only for extracting such information 
from detected gravitational waves, 
but also for a more efficient detector design 
of the presently on-going project, 
it is an urgent problem to calculate 
as precise as possible the gravitational wave signal 
expected from such a binary system. 

Because of the extreme mass ratio, 
such a binary system can be treated by a perturbation formalism. 
We treat the supermassive black hole as a background, 
and treat the solar-mass compact object as a source of metric perturbation. 
It was shown\footnote{
In the framework of metric perturbations, it is not a simple problem 
to define a point particle. 
When we take the zero-volumn limit, 
the perturbations become divergent around the particle 
and the perturbation scheme becomes invalid. 
In Ref.\cite{mino}, by using a matched asymptotic technique 
consistent with the perturbation formalism, 
we showed that the use of a point particle is still valid 
to induce a correct metric pertrubation.
} that, 
when the spatial volumn of the solar-mass compact object is 
smaller than the background curvature scale, 
one can approximately use a point source 
\begin{eqnarray}
T^{\mu\nu} \,=\, \mu \int d\tau 
{\delta^{(4)}\bigl(x-z(\tau)\bigr)\over \sqrt{-g}}
{dz^\mu\over d\tau}{dz^\nu\over d\tau}\,, \nonumber 
\end{eqnarray}
where $z^\mu(\tau)$ is the orbit of the object, 
and $\tau$ is proper time. 

By the uniqueness theorem of black holes, 
we can assume that the background black hole is described 
by the Kerr geometry. 
In this case, given an orbit, 
there is a formalism to calculate the metric perturbation 
at infinity\cite{gw}. 
Therefore, the remaining problem is 
how to calculate the orbit $z^\mu(\tau)$ of the source. 

In the massless limit, the particle moves along a geodesic. 
When the particle has a mass, 
it becomes a source of gravitational radiation, 
which carries away orbital energy and angular momentum. 
Thus the orbit deviates from a geodesic. 
The deviation from a geodesic can be derived fully 
by solving an equation of motion with the so-called self-force. 
A general calculation scheme for the gravitational self-force 
was proposed by Ref.\cite{g_self}. 
A similar situation happens to a particle 
with small scalar/electromagnetic charge, 
and the scalar and electromagnetic self-forces 
were proposed in Ref.\cite{s_e_self}. 
In this paper, 
starting from a discussion of a symmetry property of the self-force, 
we propose a method to calculate the orbital evolution 
under scalar/electromagnetic/gravitational radiation reaction. 
We assume that the effect of the radiation reaction is weak, 
and consider the leading correction to the orbital evolution. 

It is commonly assumed that the orbit evolves in an adiabatic manner, 
namely, the orbit evolves slowly in its phase space. 
A true geodesic around a Kerr black hole is characterized 
by the energy $E$, the angular momentum $L$ and the Carter constant 
$K$\footnote{We adopt 
the definition of the Carter constant in Ref.\cite{chandra}.}. 
Numerous investigations have been made 
to calculate the radiation reaction effect 
on energy $E$ and angular momentum $L$\cite{gw} 
by analyzing the asymptotic gravitational waves 
at infinity and the horizon. 
However, the calculation of the radiation reaction effect 
on the Carter constant $K$ is so far an unsolved problem. 

A geodesic equation is a set of 
second order differential equations of 4 functions $\{z^\alpha\}$. 
With a proper time $\tau$ as the parameter characterizing the orbit, 
we have 7 integral constants, 
$z^\alpha(\tau=0)$ and $dz^\alpha/d\tau(\tau=0)$, 
three of which are related to $E$, $L$ and $K$. 
In Sec.\ref{sec:sym}, we introduce a specific symmetry property 
of families of geodesics in the Kerr geometry. 
Using this symmetry, we discuss an important property of 
the self-force induced by a geodesic, 
and prove that the radiation reaction to the energy $E$, 
the angular momentum $L$ and the Carter constant $K$ 
can be derived by use of a radiative Green function 
(a half-retarded-minus-half-advanced Green function). 
However, this is not the end of the story. 
We find that the orbit does not evolve in a strictly adiabatic manner 
in general as the energy $E$, the angular momentum $L$ and 
the Carter constant $K$ vary on a short time scale. 
Besides, it is not trivial that the rest of constants do not evolve 
by the self-force. 

In order to complete the calculation of the orbit, 
we perturbatively integrate the orbital equation with the self-force 
in Sec.\ref{sec:orbit}. 
We figure out the part which secularly evolves by the self-force, 
and define the `adiabatic evolution' of the orbit 
in the way we can approximately calculate the orbital evolution. 
By showing the validity of the adiabatic approximation, 
we propose our new formalism as a conventional tool to predict 
the gravitational waves detected by the future LISA project. 

In this paper, 
we adopt Boyer-Lindquist coordinates $\{t,\,r,\,\theta,\,\phi\}$, and 
$M$, $a$ are the mass and the spin coefficient of the black hole respectively.

\section{Self-force and Symmetry} \label{sec:sym} 

The purpose of this section is to show 
a simple method to calculate the self-force. 
The core idea of this new proposal is to use a symmetry 
of the background spacetime together with the whole familiy of geodesics. 

In Subsec.\ref{sec:geo}, we first discuss 
how we can define the family of geodesics. 
Since we are interested in a particle motion 
as a target of gravitational wave observation, 
we only consider geodesics rotating around a Kerr black hole 
which neither fall into the horizon nor go to infinity. 

Subsec.\ref{sec:gps} discusses 
some symmetry properties of the Kerr spacetime. 
We apply these symmetry transformations 
to the familiy of geodesics and to the self-force induced on a geodesic 
in Subsec.\ref{sec:self}. 
Using the result of these transformation properties, 
we discuss a general expression of the evolution equations 
of the energy, angular momentum, and Carter `constant' 
in Subsec.\ref{sec:react}, 
and find that a part of the evolution equation can be evaluated 
by using the radiative Green function, 
which has a great computational advantage. 

In Subsec.\ref{sec:other}, we give some comments about practical issues 
seriously discussed in the self-force problem.

\subsection{Geodesics around a Kerr black hole} \label{sec:geo} 

A general geodesic satisfies the equations 
\begin{eqnarray}
\left({dr \over d\lambda}\right)^2
&=& [(r^2+a^2)E-aL]^2-\Delta(r^2+K) 
\,, \label{eq:g_r} \\ 
\left({d\theta\over d\lambda}\right)^2 
&=& -(aE\sin\theta-L\cosec\theta)^2 -a^2\cos^2\theta+K 
\,, \label{eq:g_the} \\ 
{dt \over d\lambda}
&=& {1\over \Delta}\left(\Sigma^2 E -2aMr L\right) \,, \label{eq:g_t} \\ 
{d\phi \over d\lambda}
&=& {1\over \Delta}\left[2aMr E +(\rho^2-2Mr) L \cosec^2\theta\right] 
\label{eq:g_p} \,, \\ 
\rho^2 &=& r^2+a^2\cos^2\theta 
\,, \quad  
\Delta \,=\, r^2-2Mr+a^2 
\,, \\ 
\Sigma^2 &=& (r^2+a^2)\rho^2+2a^2Mr\sin^2\theta 
\,. 
\end{eqnarray}
Here we use $\lambda$ as an orbital parameter 
related with the proper time $\tau$ 
by $\tau = \int^\lambda_0 d\lambda \rho^2$. 

We consider the case that the radial motion is bounded in $r_1 < r < r_2$. 
One can integrate (\ref{eq:g_r}), 
and we have a radial periodic solution 
with respect to $\lambda$ with the period 
\begin{eqnarray}
\lambda_r &=& 2\int^{r_2}_{r_1}dr 
{1 \over \sqrt{[(r^2+a^2)E-aL]^2-\Delta(r^2+K)}} \,. 
\end{eqnarray}
We write the periodic solution of the radial equation as 
\begin{eqnarray}
r(\lambda) \,=\, R(E,L,K;\lambda-\bar\lambda_r) 
\,=\, R(E,L,K;\lambda-\bar\lambda_r+\lambda_r) \,, \label{eq:g_r_f} 
\end{eqnarray}
where we set $R(E,L,K;0)=r_1$ 
(or, equivalently, $R(E,L,K;\lambda_r/2)=r_2$). 
$\bar\lambda_r$ is the integral constant of 
the first differential equation (\ref{eq:g_r}), 
and we have a periodicity in $\bar\lambda_r$, namely, 
$\bar\lambda_r \to \bar\lambda_r+n\lambda_r$ does not change the orbit, 
where $n$ is an arbitrary integer. 
Because (\ref{eq:g_r}) is invariant under $\lambda \to -\lambda$, 
$R(E,L,K;\lambda)$ becomes 
a symmetric function with respect to $\lambda$ as 
\begin{eqnarray}
R(E,L,K;\lambda) &=& R(E,L,K;-\lambda) \,. \label{eq:g_r_sym} 
\end{eqnarray}

When the orbit is radially bounded, 
the $\theta$ motion oscillates symmetrically around $\theta=\pi/2$ 
in a domain $0 < \theta_1 < \theta < \pi-\theta_1 < \pi$\cite{chandra}. 
Similarily to the radial motion, 
we have a $\theta$ periodic solution with respect to $\lambda$ 
by the period 
\begin{eqnarray}
\lambda_\theta &=& 4\int^{\pi/2}_{\theta_1}d\theta 
{1 \over \sqrt{-(aE\sin\theta-L\cosec\theta)^2 -a^2\cos^2\theta+K}} \,. 
\end{eqnarray}
We define the solution of (\ref{eq:g_the}) as 
\begin{eqnarray}
\theta(\lambda) \,=\, \Theta(E,L,K;\lambda-\bar\lambda_\theta) 
\,=\, \Theta(E,L,K;\lambda-\bar\lambda_\theta+\lambda_\theta) 
\,, \label{eq:g_the_f} 
\end{eqnarray}
where we set $\Theta(E,L,K;0)=\theta_1$ 
(or, equivalently, $\Theta(E,L,K;\lambda_\theta/2)=\pi-\theta_1$). 
$\bar\lambda_\theta$ is the integral constant in solving (\ref{eq:g_the}), 
and we have a periodicity in $\bar\lambda_\theta$ 
as in the radial equation (\ref{eq:g_r_f}). 
The symmetry property of (\ref{eq:g_the_f}) becomes 
\begin{eqnarray}
\Theta(E,L,K;\lambda) &=& \Theta(E,L,K;-\lambda) 
\,. \label{eq:g_the_sym} 
\end{eqnarray}

We write the solutions of (\ref{eq:g_t}) and (\ref{eq:g_p}) as 
\begin{eqnarray}
t(\lambda) &=& T(E,L,K,\bar\lambda_r,\bar\lambda_\theta;\lambda)
+\bar t \,, \label{eq:g_t_f} \\ 
T(E,L,K,\bar\lambda_r,\bar\lambda_\theta;\lambda) &=& \int^\lambda_0 
{d\lambda\over \Delta} \left(\Sigma^2 E -2aMr L\right) \,, \\
\phi(\lambda) &=& \Phi(E,L,K,\bar\lambda_r,\bar\lambda_\theta;\lambda)
+\bar\phi \,, \label{eq:g_p_f} \\ 
\Phi(E,L,K,\bar\lambda_r,\bar\lambda_\theta;\lambda) &=& \int^\lambda_0 
{d\lambda\over \Delta}\left[2aMr E +(\rho^2-2Mr) L \cosec^2\theta\right] 
\,, 
\end{eqnarray}
where $\bar t$ and $\bar\phi$ are the integral constants. 
We have addition formulae of 
$T(E,L,K,\bar\lambda_r,\bar\lambda_\theta;\lambda)$ 
and $\Phi(E,L,K,\bar\lambda_r,\bar\lambda_\theta;\lambda)$ as 
\begin{eqnarray}
T(E,L,K,\bar\lambda_r,\bar\lambda_\theta;\lambda) 
&=& T(E,L,K,\bar\lambda_r-\lambda_x,\bar\lambda_\theta-\lambda_x;\lambda-\lambda_x)
+T(E,L,K,\bar\lambda_r,\bar\lambda_\theta;\lambda_x) 
\,, \label{eq:g_t_sym} \\
\Phi(E,L,K,\bar\lambda_r,\bar\lambda_\theta;\lambda) 
&=& \Phi(E,L,K,\bar\lambda_r-\lambda_x,\bar\lambda_\theta-\lambda_x;\lambda-\lambda_x)
+\Phi(E,L,K,\bar\lambda_r,\bar\lambda_\theta;\lambda_x) 
\,. \label{eq:g_p_sym} 
\end{eqnarray}
By using (\ref{eq:g_r_sym}) and (\ref{eq:g_the_sym}), 
the symmetry properties of (\ref{eq:g_t_f}) and (\ref{eq:g_p_f}) become 
\begin{eqnarray}
T(E,L,K,\bar\lambda_r,\bar\lambda_\theta;\lambda) \,=\, 
-T(E,L,K,-\bar\lambda_r,-\bar\lambda_\theta;-\lambda) \,,\quad 
\Phi(E,L,K,\bar\lambda_r,\bar\lambda_\theta;\lambda) 
\,=\, -\Phi(E,L,K,-\bar\lambda_r,-\bar\lambda_\theta;-\lambda) 
\,. \label{eq:g_t_p_sym} 
\end{eqnarray}

\subsection{$t$- and $\phi$-translation 
and Geodesic Preserving Symmetry} \label{sec:gps} 

As is well-known, a Kerr spacetime has 
$t$- and $\phi$-translation symmetry. 
Applying the coordinate transformation as 
\begin{eqnarray}
t\,=\,t'+t_s,\, r\,=\,r',\, \theta\,=\,\theta',\, \phi\,=\,\phi'+\phi_s 
\,, \label{eq:trn} 
\end{eqnarray}
we have a Kerr metric of the same mass and spin parameter 
with the new coordinates $\{t',\,r',\,\theta',\,\phi'\}$. 

We consider to apply this transformation to a geodesic 
with a $\lambda$-translation, $\lambda' = \lambda-\lambda_x$, 
where $\lambda_x$ is an arbitrary constant. 
Using (\ref{eq:g_t_sym}) and (\ref{eq:g_p_sym}), we have another geodesic 
in a new coordinate system as 
\begin{eqnarray}
r'(\lambda') &=& R(E,L,K;\lambda'-\bar\lambda'_r) 
\,, \quad  
\theta'(\lambda') \,=\, \Theta(E,L,K;\lambda'-\bar\lambda'_\theta) 
\,, \\ 
t'(\lambda') &=& T(E,L,K,\bar\lambda'_r,\bar\lambda'_\theta;\lambda')
+\bar t' \,, \quad 
\phi'(\lambda') \,=\, \Phi(E,L,K,\bar\lambda'_r,\bar\lambda'_\theta;\lambda')
+\bar\phi' \,, 
\end{eqnarray}
where the orbital constants, 
$\{\bar\lambda'_r,\bar\lambda'_\theta,\bar t',\bar\phi'\}$, become 
\begin{eqnarray}
\bar\lambda'_r &=& \bar\lambda_r-\lambda_x \,, \quad 
\bar\lambda'_\theta \,=\, \bar\lambda_\theta-\lambda_x \,, \\ 
\bar t' &=& \bar t+T(E,L,K,\bar\lambda_r,\bar\lambda_\theta;\lambda_x)
-t_s \,, \quad 
\bar\phi' \,=\, \bar\phi+\Phi(E,L,K,\bar\lambda_r,\bar\lambda_\theta;\lambda_x)
-\phi_s \,. 
\end{eqnarray}
One sees that the orbital constants, $E$, $L$ and $K$, 
are invarinat under this transformation. 
We can set $\bar t'$, $\bar\phi'$ and 
either $\bar\lambda'_r$ or $\bar\lambda'_\theta$ arbitrarily 
by an appropriate choice of $t_s$, $\phi_s$ and $\lambda_x$. 

One cannot set both $\bar\lambda'_r$ and $\bar\lambda'_\theta$ zero 
at the same time since we have only one constant $\lambda_x$ to fix. 
However, because of the periodicity of the radial function (\ref{eq:g_r_f}), 
one can replace $\bar\lambda_r$ and $\bar\lambda_\theta$ by numbers 
congruent to $\bar\lambda_r$ and $\bar\lambda_\theta$ 
modulo $\lambda_r$ and $\lambda_\theta$ respectively, 
i.e. $\bar\lambda_r+n_r\lambda_r$ 
and $\bar\lambda_\theta+n_\theta\lambda_\theta$ 
where $n_r$ and $n_\theta$ are arbitrary integers. 
Using this freedom, we set $\lambda_x=\bar\lambda_r-n_r\lambda_r$, 
then we obtain $\bar\lambda'_r=0$ 
and $\bar\lambda'_\theta=\bar\lambda_\theta-\bar\lambda_r
+n_r\lambda_r+n_\theta\lambda_\theta$. 
When the ratio of $\lambda_r$ and $\lambda_\theta$ is irrational, 
there is a choice of $n_r$ and $n_\theta$ 
with which $|\bar\lambda'_\theta|$ become infinitestimally small, 
and a geodesic is characterized only by $E$, $L$ and $K$. 
In the following, we assume that 
the ratio of $\lambda_r$ and $\lambda_\theta$ is irrational 
though we do not set $\bar\lambda_r$ and $\bar\lambda_\theta$ zero 
for the latter conveience unless stated.

Using this transformation property, one can prove a useful formula 
of a scalar function geometrically defined along a geodesic. 
As the geodesic is characterzed by 7 constants, 
$\{E,L,K,\bar\lambda_r,\bar\lambda_\theta,\bar t,\bar\phi\}$, 
we write the scalar function 
as $f(E,L,K,\bar\lambda_r,\bar\lambda_\theta,\bar t,\bar\phi_0;\lambda)$. 
We assume that the function is invariant under $t$- and $\phi$-translation, 
then the function is independ on $\bar t$ and $\bar\phi$. 
Since the function is periodic 
with respect to $\bar\lambda_r$ and $\bar\lambda_\theta$, 
one can expand the function with discrete Fourier series, 
$e^{-i2m\pi\bar\lambda_r/\lambda_r-i2n\pi\bar\lambda_\theta/\lambda_\theta}$. 
By applying the $\lambda$-translation with $\lambda_x=\lambda$, 
we finally have 
\begin{eqnarray}
f(E,L,K,\bar\lambda_r,\bar\lambda_\theta,\bar t,\bar\phi;\lambda) 
&=& \sum_{m,n}f^{(m,n)}(E,L,K) \exp\Biggl[i2\pi\biggl(
m{\lambda-\bar\lambda_r\over \lambda_r}
+n{\lambda-\bar\lambda_\theta\over\lambda_\theta}
\biggr)\Biggr] \,. \label{eq:f_per} 
\end{eqnarray}

We next consider the symmetry as 
\begin{eqnarray}
t\,=\,-t',\, r\,=\,r',\, \theta\,=\,\theta',\, \phi\,=\,-\phi' 
\,. \label{eq:gps} 
\end{eqnarray}
By this coordinate transformation, we recover the same line element 
with the coordinates $\{t',\,r',\,\theta',\,\phi'\}$. 

We consider the transformation of a geodesic by this symmetry. 
Since we change the time direction, we transform the orbital parameter 
as $\lambda' = -\lambda$. 
Using (\ref{eq:g_r_sym}), (\ref{eq:g_the_sym}) and (\ref{eq:g_t_p_sym}), 
a geodesic is transformed to a new geodesic as 
\begin{eqnarray}
r'(\lambda') &=& R(E,L,K;\lambda'-\bar\lambda'_r) \,, \quad 
\theta'(\lambda') \,=\, \Theta(E,L,K;\lambda'-\bar\lambda'_\theta) \,, \\ 
t'(\lambda') &=& T(E,L,K,\bar\lambda'_r,\bar\lambda'_\theta;\lambda')
+\bar t' \,, \quad 
\phi'(\lambda') \,=\, \Phi(E,L,K,\bar\lambda'_r,\bar\lambda'_\theta;\lambda')
+\bar\phi' \,, 
\end{eqnarray}
where the orbital constants, 
$\{\bar\lambda'_r,\bar\lambda'_\theta,\bar t',\bar\phi'\}$, become 
\begin{eqnarray}
\bar\lambda'_r &=& -\bar\lambda_r \,, \quad 
\bar\lambda'_\theta \,=\, -\bar\lambda_\theta \,, \\ 
\bar t' &=& -\bar t \,, \quad \bar\phi' \,=\, -\bar\phi \,. 
\end{eqnarray}
Since $E$, $L$ and $K$ are invariant, 
by using the $t$- and $\phi$- translation symmetry (\ref{eq:trn}) 
in an appropriate manner, 
the new geodesic becomes equal to the original one. 
For this reason, we call this 
Geodesic Preserving Symmetry (hereafter GPS) transformation.

\subsection{Green Function and Self-force} \label{sec:self} 

We denote a various scalar/electromagnetic/gravitational Green function by 
\begin{eqnarray}
{\cal G}(x,z) \,=\, \displaystyle\cases{
G(x,z) \,, & scalar \,, \cr
G_{\alpha\,\mu}(x,z)dx^\alpha dz^\mu \,, & electromagnetism \,, \cr
G_{\alpha\beta\,\mu\nu}(x,z)dx^\alpha dx^\beta dz^\mu dz^\nu \,, & 
linear gravity \,.}
\end{eqnarray}
and consider $t$- and $\phi$-translation 
and GPS transformation property of it.

Using a symmetric scalar/electromagnetic/gravitational 
Green function ${\cal G}^{sym.}(x,z)$, 
a retarded and advanced scalar/electromagnetic/gravitational Green function, 
${\cal G}^{ret./adv.}(x,z)$ 
can be written as 
\begin{eqnarray}
{\cal G}^{ret.}(x,z) &=& 2\theta[\Sigma(x),z]{\cal G}^{sym.}(x,z) \,, \\ 
{\cal G}^{adv.}(x,z) &=& 2\theta[z,\Sigma(x)]{\cal G}^{sym.}(x,z) \,, 
\end{eqnarray}
where $\Sigma(x)$ is an arbitrary space-like hypersurface containing $x$, 
and $\theta[\Sigma(x),z]=1-\theta[z,\Sigma(x)]$ is equal to $1$ 
when $z$ lies in the past of $\Sigma(x)$ and vanishes otherwise. 
The symmetric Green function is invariant 
under $t$- and $\phi$-translation and GPS transformation 
because, in its Hadamard construction\cite{g_self,s_e_self}, 
it is described only by geometrically defined bi-tensors invariant 
under $t$- and $\phi$-translation and GPS transformation. 

Under $t$- and $\phi$-translation (\ref{eq:trn}), 
the factor $\theta[\Sigma(x),z]$ is also invariant and we have 
\begin{eqnarray}
{\cal G}^{ret.}(x',z') \,=\, {\cal G}^{ret.}(x,z) \,, \quad 
{\cal G}^{adv.}(x',z') \,=\, {\cal G}^{adv.}(x,z) \,. 
\end{eqnarray}
On the other hand, 
GPS transformation (\ref{eq:gps}) changes the direction of the time 
and the factor $\theta[\Sigma(x),z]$ transformes as 
\begin{eqnarray}
\theta[\Sigma(x'),z'] \,=\, \theta[z,\Sigma(x)] \,, \quad 
\theta[z',\Sigma(x')] \,=\, \theta[\Sigma(x),z] \,. 
\end{eqnarray}
Thence, by GPS transformation (\ref{eq:gps}), 
a retarded and advanced Green function are transformed 
to be an advanced and retarded Green function respectively as 
\begin{eqnarray}
{\cal G}^{ret.}(x',z') \,=\, {\cal G}^{adv.}(x,z) \,, \quad 
{\cal G}^{adv.}(x',z') \,=\, {\cal G}^{ret.}(x,z) \,. 
\end{eqnarray}

We next consider 
the scalar/electromagnetic/gravitational self-force acting on the particle. 
Because the field induced by a point particle diverges along the orbit, 
we need a regularization calculation to derive the self-force
\cite{g_self,s_e_self}. 
Based on the Green function method in calculating the field, 
an elegant method of regularization was proposed\cite{g_reg}, 
in which the self-force can be directly derived from the field 
calculated by the so-called R-part of a retarded Green function. 
The R-part of a retarded and advanced Green function 
${\cal G}^{R-ret./R-adv.}(x,z)$ is schematically defined as 
\begin{eqnarray}
{\cal G}^{R-ret./R-adv.}(x,z) \,=\, 
{\cal G}^{ret./adv.}(x,z) -{\cal G}^{S}(x,z) \,, 
\end{eqnarray}
where ${\cal G}^{S}(x,z)$ is the so-called S-part\cite{g_reg}. 
It is important to note that 
the R-part of the half-retarded-minus-half-advance Green function 
becomes a radiative Green function ${\cal G}^{rad.}(x,z)$ as 
\begin{eqnarray}
{1\over 2}\left({\cal G}^{R-ret.}(x,z)-{\cal G}^{R-adv.}(x,z)\right) 
\,=\, {1\over 2}\left({\cal G}^{ret.}(x,z)-{\cal G}^{adv.}(x,z)\right) 
\,=\, {\cal G}^{rad.}(x,z) \,. \label{eq:rad} 
\end{eqnarray}

Similar to the symmetric Green function, 
the S-part is defined by geometric bi-tensors 
and is invariant both by $t$- and $\phi$-translation and GPS transformation. 
Thence the R-part of the retarded and advanced Green function
are still invariant under $t$- and $\phi$-translation, 
and, by GPS transformation, the R-part becomes 
\begin{eqnarray}
{\cal G}^{R-ret.}(x',z') \,=\, {\cal G}^{R-adv.}(x,z) \,, \quad 
{\cal G}^{R-adv.}(x',z') \,=\, {\cal G}^{R-ret.}(x,z) \,. 
\label{eq:gps_g} 
\end{eqnarray}

The scalar/electromagnetic/gravitational self-force 
is schemetically described as 
\begin{eqnarray}
F^{R-ret./R-adv.}_\alpha(\tau) 
&=& \lim_{x\to z(\tau)}F_\alpha[\phi^{R-ret./R-adv.}](x) \,, \\ 
\phi^{R-ret./R-adv.}(x) 
&=& \int d\tau G^{R-ret./R-adv.}(x,z(\tau))S(z(\tau)) \,, 
\end{eqnarray}
where $\phi^{R-ret./R-adv.}$ is 
the R-part of a scalar/electromagnetic/gravitational potential 
using the R-part of a retarded/advanced Green function, 
and we note $^{R-ret./R-adv.}$ to the self-force to emphasize that 
it is derived using the R-part of the retarded/advanced Green function. 
$S(z(\tau))$ is the source term defined along the orbit. 
We assume that the tensor differential operator $F_\alpha[](x)$ is defined 
to satisfy the normalization condition as 
$F_\alpha[](z(\tau))v^\alpha(\tau)=0$. 

We assume that the self-force is weak 
and the orbit can be approximated to be a geodesic at each instant of time. 
Using this approximation, we consider to calculate the self-force 
induced by a geodesic. 
We write the self-force as a vector function of the orbital constants 
$E$, $L$, $K$, $\bar\lambda_r$, $\bar\lambda_\theta$, $\bar t$, $\bar\phi$ 
and the orbital parameter $\lambda$ as 
\begin{eqnarray}
F^{R-ret./R-adv.}_\alpha &=& 
F^{R-ret./R-adv.}_\alpha
(E,L,K,\bar\lambda_r,\bar\lambda_\theta,\bar t,\bar\phi;\lambda) 
\,. \label{eq:self_f_geo} 
\end{eqnarray}
In general, 4-velocity and a self-force transform as 
\begin{eqnarray}
v^{\alpha'} &=& {dx'^\alpha \over d\tau'}(\tau') 
\,=\, \left({\partial x'^\alpha\over\partial x^\beta}\right)
\left({d\tau\over d\tau'}\right)v^\beta(\tau) \,, \\ 
f_{\alpha'} &=& {D \over d\tau'}v_{\alpha'}(\tau') 
\,=\, \left({d\tau\over d\tau'}\right){D\over d\tau}\left[
\left({\partial x^\beta\over\partial x'^\alpha}\right)
\left({d\tau\over d\tau'}\right)v_\beta\right](\tau) \,. 
\end{eqnarray} 
Using these transformation rules, 
we apply $t$- and $\phi$-translation (\ref{eq:trn}) 
and GPS transformation (\ref{eq:gps}) 
to the self-force induced by a geodesic (\ref{eq:self_f_geo}). 

We first consider $t$- and $\phi$-translation (\ref{eq:trn}) 
with $\lambda_x=0$. 
Applying the coordinate transformation, we have 
\begin{eqnarray}
F^{R-ret./R-adv.}_{\alpha'}
(E,L,K,\bar\lambda_r,\bar\lambda_\theta,\bar t+t_s,\bar\phi+\phi_s;\lambda) 
&=& F^{R-ret./R-adv.}_\alpha
(E,L,K,\bar\lambda_r,\bar\lambda_\theta,\bar t,\bar\phi;\lambda) \,. 
\end{eqnarray}
Since the metric is invariant under the transformation, we have 
$F^{R-ret./R-adv.}_{\alpha'}=F^{R-ret./R-adv.}_\alpha$, 
thus, the self-force does not depend on $\bar t$ and $\bar\phi$. 
In the following, we write the self-force as 
$F^{R-ret./R-adv.}_\alpha(E,L,K,\bar\lambda_r,\bar\lambda_\theta;\lambda)$. 

Finally we consider GPS transformation of the self-force. 
Noting (\ref{eq:gps_g}), (\ref{eq:gps}) transforms the self-force as 
\begin{eqnarray}
F^{R-adv.}_\alpha(E,L,K,-\bar\lambda_r,-\bar\lambda_\theta;-\lambda) 
&=& (-1)^s F^{R-ret.}_\alpha(E,L,K,\bar\lambda_r,\bar\lambda_\theta;\lambda) 
\,, \\ 
F^{R-ret.}_\alpha(E,L,K,-\bar\lambda_r,-\bar\lambda_\theta;-\lambda) 
&=& (-1)^s F^{R-adv.}_\alpha(E,L,K,\bar\lambda_r,\bar\lambda_\theta;\lambda) 
\,, \label{eq:gps_f} 
\end{eqnarray}
where $s=1$ for $\alpha=t,\phi$, and $s=0$ for $\alpha=r,\theta$.

\subsection{Evolution of the energy, angular momentum and Carter `constant'} 
\label{sec:react} 

As we use $\lambda$ as an orbital parameter, 
we consider the $\lambda$ derivative of these `constants' as 
\begin{eqnarray}
\left[{d\over d\lambda}E\right]^{R-ret./R-adv.}
&=& -\rho^2F^{R-ret./R-adv.}_t
(E,L,K,\bar\lambda_r,\bar\lambda_\theta;\lambda) 
\,, \label{eq:g_E_f} \\
\left[{d\over d\lambda}L\right]^{R-ret./R-adv.}
&=& \rho^2F^{R-ret./R-adv.}_\phi
(E,L,K,\bar\lambda_r,\bar\lambda_\theta;\lambda) 
\,, \label{eq:g_L_f} \\  
\left[{d\over d\lambda}K\right]^{R-ret./R-adv.}
&=& \Bigl[ 2(r^2+a^2)\rho^2v^tF^{R-ret./R-adv.}_t
+2\rho^4v^rF^{R-ret./R-adv.}_r \Bigr]
(E,L,K,\bar\lambda_r,\bar\lambda_\theta;\lambda) 
\,. \label{eq:g_K_f} 
\end{eqnarray}
Since these are scalar functions defined along a geodesic, 
we can apply the formula (\ref{eq:f_per}) 
to (\ref{eq:g_E_f}), (\ref{eq:g_L_f}) and (\ref{eq:g_K_f}), 
and we obtain 
\begin{eqnarray}
\left[{d\over d\lambda}{\cal E}\right]^{R-ret./R-adv.}
&=& \sum_{m,n} \dot {\cal E}^{R-ret./R-adv.(m,n)}(E,L,K) 
\exp\biggl[i2\pi\Biggl(m{\lambda-\bar\lambda_r\over\lambda_r}
+n{\lambda-\bar\lambda_\theta\over\lambda_\theta}\biggr)\Biggr] 
\,, \label{eq:g_per} 
\end{eqnarray}
where we denote $E$, $L$ and $K$ by ${\cal E}$. 
By the reality condition, we have 
$(\dot {\cal E}^{R-ret./R-adv.\,(m,n)})^*
=\dot {\cal E}^{R-ret./R-adv.\,(-m,-n)}$, 
where ${}^*$ means we take the complex conjugation operation. 

4-velocity of a geodesic transforms by (\ref{eq:gps}) as 
\begin{eqnarray}
v^{\alpha'}(E,L,K,-\bar\lambda_r,-\bar\lambda_\theta;-\lambda) 
&=& (-1)^s v^\alpha(E,L,K,\bar\lambda_r,\bar\lambda_\theta;\lambda) 
\,,\label{eq:gps_v} 
\end{eqnarray}
where $s=0$ for $\alpha=t,\phi$, and $s=1$ for $\alpha=r,\theta$. 
Using (\ref{eq:gps_f}) and (\ref{eq:gps_v}), 
(\ref{eq:g_E_f}), (\ref{eq:g_L_f}) and (\ref{eq:g_K_f}) transform as 
\begin{eqnarray}
\left[{d\over d\lambda}{\cal E}\right]^{R-adv./R-ret.}
(E,L,K,-\bar\lambda_r,-\bar\lambda_\theta;-\lambda) 
&=& -\left[{d\over d\lambda}{\cal E}\right]^{R-ret./R-adv.}
(E,L,K,\bar\lambda_r,\bar\lambda_\theta;\lambda) 
\,. \label{eq:gps_e} 
\end{eqnarray}

We consider the evolution euations, 
(\ref{eq:g_E_f}), (\ref{eq:g_L_f}) and (\ref{eq:g_K_f}), 
averaged at two orbital points characterized as 
$z(E,L,K,\bar\lambda_r,\bar\lambda_\theta,\bar t,\bar\phi;\lambda)$ 
and $z(E,L,K,-\bar\lambda_r,-\bar\lambda_\theta,\bar t',\bar\phi';-\lambda)$. 
Using (\ref{eq:gps_e}), one finds 
the evolution equations are described 
by the radiative Green function (\ref{eq:rad}) 
instead of the R-part of the retarded/advanced Green function as 
\begin{eqnarray}
&& {1\over 2}\Biggl\{\biggl[{d\over d\lambda}{\cal E}\biggr]^{R-ret.}
(E,L,K,\bar\lambda_r,\bar\lambda_\theta;\lambda) 
+\biggl[{d\over d\lambda}{\cal E}\biggr]^{R-ret.}
(E,L,K,-\bar\lambda_r,-\bar\lambda_\theta;-\lambda)\Biggr\}
\nonumber \\ && \qquad \qquad 
\,=\, -{1\over 2}\Biggl\{\biggl[{d\over d\lambda}{\cal E}\biggr]^{R-adv.}
(E,L,K,\bar\lambda_r,\bar\lambda_\theta;\lambda) 
+\biggl[{d\over d\lambda}{\cal E}\biggr]^{R-adv.}
(E,L,K,-\bar\lambda_r,-\bar\lambda_\theta;-\lambda)\Biggr\} 
\nonumber \\ && \qquad \qquad 
\,=\, \Biggl[{d\over d\lambda}{\cal E}\Biggr]^{rad.}
(E,L,K,\bar\lambda_r,\bar\lambda_\theta;\lambda) 
\,, \\ && \qquad 
\Biggl[{d\over d\lambda}E\Biggr]^{rad.}
(E,L,K,\bar\lambda_r,\bar\lambda_\theta;\lambda)
\,=\, -\rho^2 F^{rad.}_t(E,L,K,\bar\lambda_r,\bar\lambda_\theta;\lambda) 
\,, \\ && \qquad 
\Biggl[{d\over d\lambda}L\Biggr]^{rad.}
(E,L,K,\bar\lambda_r,\bar\lambda_\theta;\lambda)
\,=\, \rho^2 F^{rad.}_\phi(E,L,K,\bar\lambda_r,\bar\lambda_\theta;\lambda) 
\,, \\ && \qquad 
\Biggl[{d\over d\lambda}K\Biggr]^{rad.}
(E,L,K,\bar\lambda_r,\bar\lambda_\theta;\lambda)
\,=\, \Bigl[ 2(r^2+a^2)\rho^2v^tF^{rad.}_t
+2\rho^4v^rF^{rad.}_r \Bigr](E,L,K,\bar\lambda_r,\bar\lambda_\theta;\lambda) 
\,, 
\end{eqnarray} 
where $F^{rad.}_\alpha$ is the self-force 
calculated by the radiative Green function (\ref{eq:rad}). 

We write 
\begin{eqnarray}
\biggl[{d\over d\lambda}{\cal E}\biggr]^{rad.}
(E,L,K,\bar\lambda_r,\bar\lambda_\theta;\lambda) 
&=& \sum_{m,n} \dot {\cal E}^{rad.\,(m,n)}(E,L,K) 
\exp\biggl[i2\pi\Biggl(m{\lambda-\bar\lambda_r\over\lambda_r}
+n{\lambda-\bar\lambda_\theta\over\lambda_\theta}\biggr)\Biggr] 
\,. 
\end{eqnarray}
then we have 
\begin{eqnarray}
{1\over 2}\bigl(\dot{\cal E}^{R-ret.\,(m,n)}
+\dot{\cal E}^{R-ret.\,(-m,-n)}\bigl)(E,L,K) 
&=& -{1\over 2}\bigl(\dot{\cal E}^{R-adv.\,(m,n)}
+\dot{\cal E}^{R-adv.\,(-m,-n)}\bigl)(E,L,K) 
\nonumber \\ 
&=& \dot{\cal E}^{rad.\,(m,n)}(E,K,L) \,. \label{eq:part} 
\end{eqnarray} 
Thus, half of the expansion coefficients of the evolution equations 
can be derived by using the radiative Green function. 

We comment that, 
when the ratio of $\lambda_r$ and $\lambda_\theta$ is irrational, 
our formula genelarizes the result in Ref.\cite{galtsov}, 
in which it is proven that 
radiation reaction to the energy and the angular momentum 
along a whole geodesic 
can be derived by a self-force 
calculated by a radiative Green function\footnote{
We also note that our formalism specifies the case that 
the orbit inducing the self-force can approximated by a geodesic, 
while the formula in Ref.\cite{galtsov} applies to a general orbit 
in scalar/electromagnetic case.}. 
By (\ref{eq:g_per}), the radiation reaction averaged per unit $\lambda$ 
to the energy, angular momentum and Carter `constant' becomes 
\begin{eqnarray}
\lim_{\lambda\to\infty}{1\over 2\lambda}\int^\lambda_{-\lambda}d\lambda
{d\over d\lambda}{\cal E}^{R-ret.}
(E,L,K,\bar\lambda_r,\bar\lambda_\theta;\lambda) 
&=& \dot{\cal E}^{R-ret.\,(0,0)}(E,L,K) \,. \label{eq:react0}
\end{eqnarray}
By (\ref{eq:part}), (\ref{eq:react0}) agrees 
with the calculation using the radiative Green function as 
\begin{eqnarray}
\lim_{\lambda\to\infty}{1\over 2\lambda}\int^\lambda_{-\lambda}d\lambda
{d\over d\lambda}{\cal E}^{rad.}
(E,L,K,\bar\lambda_r,\bar\lambda_\theta;\lambda) 
&=& \dot{\cal E}^{R-ret.\,(0,0)}(E,L,K) 
\,. \label{eq:galtsov}
\end{eqnarray}
It is notable that 
the dependence of both $\bar\lambda_r$ and $\bar\lambda_\theta$ 
vanishes in the end.

\subsection{Practical issues in calculating the radiative potential} 
\label{sec:other}

The primary problem in a Kerr case is 
we have no conventional method to calculate 
a electromagnetic potential and a metric perturbation 
induced by a point particle
\footnote{Recently, some ideas to calculate 
the vector potential and the metric perturbation induced by a point source 
were proposed\cite{kerr}.}. 
The construction of an inhomogeneous solution is unknown in general, 
however, a very simple method to derive homogeneous solutions 
was proposed\cite{chrz}. 
In Ref.\cite{chrz}, it was also discussed 
the derivation of the retarded and advanced Green functions 
as infinite sums of homogeneous solutions, 
which gives a correct metric perturbation only outside the source. 
For example, when the particle moves in the radial domain $r_{min}<r<r_{max}$, 
the metric perturbation given in Ref.\cite{chrz} is correct 
at $r>r_{max}$ and $r<r_{min}$. 

Though the prescription in Ref.\cite{chrz} is insufficient 
for inhomogeneous Green functions, 
it gives the correct radiative Green function 
since it is just a sum of homogeneous solutions. 
Suppose we calculate the radiative Green function following Ref.\cite{chrz}, 
it is correct outside the source. 
However, since it is made 
as an infinite sum of homogeneous solutions by construction, 
it satisfies the source-free Einstein equations 
at every radial domain. 
Thus, it is a correct radiative Green function in the whole spacetime.

\section{Perturbative Evolution of an orbit} \label{sec:orbit} 

To make a definite discussion, we consider that, 
at $\lambda<0$, the particle moves along a geodesic 
characterized by the constants, ${\cal E}={\cal E}_0$, 
$\bar\lambda_r=\bar\lambda_{r0}$, 
$\bar\lambda_\theta=\bar\lambda_{\theta 0}$, 
$\bar t=\bar t_0$ and $\bar\phi=\bar\phi_0$, 
and that the self-force begins to act on the orbit when $\lambda>0$, 
and deviate from the initial geodesic. 
In this section, we discuss 
the deviation of the initial geodesic in a perturbative manner. 
We define $\mu$ as the charge or the mass of the orbiting particle 
normalized by the mass of the background black hole, 
and we consider $\mu$ is an infinitestimally small value 
as an index of the perturbation. 

In order to see how the orbit evolves by the self-force, 
we first consider (\ref{eq:g_per}). 
We define the deviation of ${\cal E}$ 
from the initial value ${\cal E}_0$ by $\delta{\cal E}$. 
Because we only consider the self-force induced by a geodesic 
in deriving (\ref{eq:g_per}), 
we can consistenly derive the evolution of $\delta{\cal E}$ 
only when $\delta{\cal E}=O(\mu^\alpha),\,\alpha>0$. 
The evolution of $\delta{\cal E}$ becomes 
\begin{eqnarray}
\delta{\cal E}(E_0,L_0,K_0,\bar\lambda_{r0},\bar\lambda_{\theta 0};\lambda)
&=& \lambda \dot{\cal E}^{R-ret.}(E_0,L_0,K_0)
+\sum_{m,n} {\cal E}^{R-ret.\,(m,n)}(E_0,L_0,K_0) 
e^{i2\pi\bigl(m{\lambda-\bar\lambda_{r0}\over\lambda_r}
+n{\lambda-\bar\lambda_{\theta 0}\over\lambda_\theta}\bigr)}
\,, \label{eq:e_evo} 
\end{eqnarray}
where the coefficient of the linearly growing term is determined as 
\begin{eqnarray}
\dot{\cal E}^{R-ret.} \,=\, \dot{\cal E}^{R-ret.\,(0,0)} \,. 
\end{eqnarray}
One sees that (\ref{eq:e_evo}) consists of two parts; 
the secular part and the oscillating part. 
The oscillating part stays $O(\mu^1)$ at any $\lambda$. 
On the other hand, the secular part grows linearly by $\lambda$, 
thus, one can consistently derive $\delta{\cal E}$ 
only when $\lambda$ is $O(\mu^\alpha),\,0\geq\alpha>-1$. 

Because of this oscillating term, 
one cannot say that the orbit evolves adiabatically in an exact sense. 
The oscillating part shows the interaction of 
the orbit and the `heat bath' of radiation. 
In the time scale of the order $O(\mu^0)$, 
the orbit just exchanges the energy, angular momentum with the `heat bath' 
and they increase and decrease in the equal rate. 
In the long time scale of the order $O(\mu^\alpha),\,0>\alpha>-1$, 
the energy and angular momentum reserved in the `heat bath' escape 
into the horizon or away to infinity 
and the orbital energy and angular momentum 
tend to flow out to the `heat bath'. 
Thus, as described by the secular part of (\ref{eq:e_evo}), 
the orbital energy and angular momentum decrease linearly by $\lambda$. 

Though we do not have an adiabatic evolution in an exact sense, 
we show that the secular part of the `constants' ${\cal E}$ 
becomes dominant over the oscilating part. 
If the same thing happens to the rest of `constants', 
it seems possible to define an `adiabatic' evolution of the orbit 
in an approximate sense. 
For this purpose, 
we discuss an orbital evolution in a perturbative manner. 
We first consider the evolution of 
$r$ and $\theta$ coordinates in Subsec.\ref{sec:r-the}, 
then, $t$ and $\phi$ coordinates in Subsec.\ref{sec:t-p}. 
Subsec.\ref{sec:adi} gives 
a plausible definition of an `adiabatic' evolution of the orbit, 
which approximates the exact orbital evolution by a self-force. 
Subsec.\ref{sec:gauge} concludes the section 
with a discussion of a gauge dependence of an `adiabatic' evolution 
which appears only in gravitational case. 

We define an orbit evolving by a self-force as 
\begin{eqnarray}
t(\lambda) &=& t_0(\lambda)+\delta t(\lambda) \,, \quad 
r(\lambda) \,=\, r_0(\lambda)+\delta r(\lambda) \,, \\ 
\theta(\lambda) &=& \theta_0(\lambda)+\delta\theta(\lambda) \,, \quad 
\phi(\lambda) \,=\, \phi_0(\lambda)+\delta\phi(\lambda) \,, 
\end{eqnarray}
where $\{t_0,r_0,\theta_0,\phi_0\}$ is the initial geodesic. 
For the latter convenience, we define a family of geodesics as 
\begin{eqnarray}
t(\lambda) &=& T(E,L,K,\bar\lambda_r,\bar\lambda_\theta;\lambda)
+\bar t \,, \quad 
r(\lambda) \,=\, R(E,L,K;\lambda-\bar\lambda_r) \,, \\ 
\theta(\lambda) &=& \Theta(E,L,K;\lambda-\bar\lambda_r) \,, \quad 
\phi(\lambda) \,=\, \Phi(E,L,K,\bar\lambda_r,\bar\lambda_\theta;\lambda)
+\bar \phi \,, 
\end{eqnarray}

\subsection{$r$-motion and $\theta$-motion} \label{sec:r-the} 

Instead of integrating the equation of motion $Dv^\alpha/d\tau=F^\alpha$, 
we consider to integrate (\ref{eq:g_r}) and (\ref{eq:g_the}) 
to derive the motion of $r$ and $\theta$ coodinates. 
For a convenience, we write (\ref{eq:g_r}) and (\ref{eq:g_the}) as 
\begin{eqnarray}
\left({dr \over d\lambda}\right)^2 \,=\, V({\cal E},r) \,, \quad 
\left({d\theta\over d\lambda}\right)^2 \,=\, U({\cal E},\theta) 
\,, \label{eq:r_the} 
\end{eqnarray}
where ${\cal E}={\cal E}_0+\delta{\cal E}$, 
$r=r_0+\delta r$ and $\theta=\theta_0+\delta\theta$. 

Taking the leading order deviation from the initial geodesic, 
(\ref{eq:r_the}) become 
\begin{eqnarray} 
2{d\delta r \over d\lambda}{dr_0 \over d\lambda} &=& 
V({\cal E}_0,r_0)_{,i}\delta {\cal E}^i
+V({\cal E}_0,r_0)_{,r}\delta r 
\,, \quad 
2{d\delta\theta\over d\lambda}{d\theta_0 \over d\lambda} \,=\, 
U({\cal E}_0,\theta_0)_{,i}\delta {\cal E}^i
+U({\cal E}_0,r_0)_{,\theta}\delta\theta 
\,, \label{eq:d_r_the0}
\end{eqnarray} 
where we denote 
$f_{,i}\delta {\cal E}^i=f_{,E}\delta E+f_{,L}\delta L+f_{,K}\delta K$. 
Using $2d^2r_0/d\lambda^2=V({\cal E}_0,r_0)_{,r}$ 
and $2d^2\theta_0/d\lambda^2=U({\cal E}_0,\theta_0)_{,\theta}$, 
we have 
\begin{eqnarray} 
{d\over d\lambda}\left({\delta r\over{dr_0\over d\lambda}}\right) 
\,=\, {1\over 2}{V_{,i}\over V}({\cal E}_0,r_0)\delta {\cal E}^i 
\,, \quad 
{d\over d\lambda}\left({\delta\theta\over{d\theta_0\over d\lambda}}\right) 
\,=\, {1\over 2}{U_{,i}\over U}({\cal E}_0,\theta_0)\delta {\cal E}^i 
\,. \label{eq:d_r_the} 
\end{eqnarray} 
The differential equations (\ref{eq:d_r_the}) have singularies 
because $dr_0/d\lambda$ and $V({\cal E}_0,r_0)$ vanish 
at $\lambda=\bar\lambda_r+(n/2)\lambda_r$, 
and $d\theta_0/d\lambda$ and $U({\cal E}_0,\theta_0)$ vanish 
at $\lambda=\bar\lambda_\theta+(n/2)\lambda_\theta$, 
where $n$ is an integer. 

One must integrate (\ref{eq:d_r_the}) 
such that $\delta r$ and $\delta\theta$ are smooth 
at the singularities. 
We formally integrate the differential equations as 
\begin{eqnarray} 
\delta r 
&=& {dr_0\over d\lambda}\bar V_i\delta {\cal E}^i 
-{dr_0\over d\lambda}\int^\lambda_0 d\lambda 
\bar V_i{d\over d\lambda}\delta {\cal E}^i 
+c_u^{(n)}{dr_0\over d\lambda} \,, \label{eq:d_r0} \\ 
\delta\theta 
&=& {d\theta_0\over d\lambda}\bar U_i\delta {\cal E}^i 
-{d\theta_0\over d\lambda}\int^\lambda_0 d\lambda 
\bar U_i{d\over d\lambda}\delta {\cal E}^i 
+c_v^{(n)}{d\theta_0\over d\lambda} \,, \label{eq:d_the0} 
\end{eqnarray} 
where we define $d\bar V_i/d\lambda = V_{,i}/2V$ 
and $d\bar U_i/d\lambda = U_{,i}/2U$. 
Here one must add the integration constants $c_u^{(n)}$ 
at $\bar\lambda_r+(n+1)\lambda_r/2>\lambda>\bar\lambda_r+n\lambda_r/2$, 
and $c_v^{(n)}$ at $\bar\lambda_\theta+(n+1)\lambda_\theta/2>
\lambda>\bar\lambda_\theta+n\lambda_\theta/2$, 
independently for each integer $n$, 
such that $\delta r=0$ and $\delta\theta=0$ at $\lambda=0$. 
and $\delta r$ and $\delta\theta$ become smooth 
at the singularities of (\ref{eq:d_r_the}). 

In order to determine $c_v^{(n)}$ and $v_u^{(n)}$ 
together with $\bar V_i$ and $\bar U_i$, 
we consider the singular structure of (\ref{eq:d_r_the}). 
We write 
\begin{eqnarray}
V({\cal E}_0,r) &=& v({\cal E}_0,r)
\Bigl(r-r_1({\cal E}_0)\Bigr)\Bigl(r_2({\cal E}_0)-r\Bigr) 
\,, \label{eq:vv} \\ 
U({\cal E}_0,\theta) &=& u({\cal E}_0,\theta)
\Bigl(\theta-\theta_1({\cal E}_0)\Bigr)
\Bigl(\pi-\theta_1({\cal E}_0)-\theta\Bigr) 
\,, \label{eq:uu}
\end{eqnarray}
where $v({\cal E}_0,r)$ is positive at $r_1<r<r_2$, 
and $u({\cal E}_0,\theta)$ is positive at $\theta_1<\theta<\pi-\theta_1$. 
$r_0$ and $\theta_0$ of the initial geodesic around the singularities 
behave as 
\begin{eqnarray}
r_0 &=& \displaystyle\cases{
r_1+{1\over4}v_1(r_2-r_1)(\lambda-\bar\lambda_r-n\lambda_r)^2
+O\Bigl((\lambda-\bar\lambda_r-n\lambda_r)^4\Bigr) \,, \cr 
r_2-{1\over4}v_2(r_2-r_1)
(\lambda-\bar\lambda_r-(n+/2)\lambda_r)^2
+O\Bigl((\lambda-\bar\lambda_r-(n+1/2)\lambda_r)^4\Bigr) \,.} 
\\ 
\theta_0 &=& \displaystyle\cases{
\theta_1+{1\over4}u_0(\pi-2\theta_1)
(\lambda-\bar\lambda_\theta-n\lambda_\theta)^2
+O\Bigl((\lambda-\bar\lambda_\theta-n\lambda_\theta)^4\Bigr) \,, \cr 
\pi-\theta_1-{1\over4}u_0(\pi-2\theta_1)
(\lambda-\bar\lambda_\theta-(n+/2)\lambda_\theta)^2
+O\Bigl((\lambda-\bar\lambda_\theta-(n+1/2)\lambda_\theta)^4\Bigr) \,.} 
\end{eqnarray}
where $v_1=v({\cal E}_0,r_1)$, $v_2=v({\cal E}_0,r_2)$, 
$u_0=u({\cal E}_0,\theta_1)=u({\cal E}_0,\pi-\theta_1)$, 
and $n$ is an integer. 
Thus, the singular structure of ${1\over 2}{V_{,i}\over V}({\cal E}_0,r_0)$ 
and ${1\over 2}{U_{,i}\over U}({\cal E}_0,\theta_0)$ becomes 
\begin{eqnarray}
{1\over 2}{V_{,i}\over V}({\cal E}_0,r_0)
&=& \displaystyle\cases{
-{2r_{1,i}\over v_1(r_2-r_1)}
(\lambda-\bar\lambda_r-n\lambda_r)^{-2}
+O\Bigl((\lambda-\bar\lambda_r-n\lambda_r)^0\Bigr) \,, \cr 
{2r_{2,i}\over v_2(r_2-r_1)}
(\lambda-\bar\lambda_r-(n+/2)\lambda_r)^{-2}
+O\Bigl((\lambda-\bar\lambda_r-(n+1/2)\lambda_r)^0\Bigr) \,,} 
\\
{1\over 2}{U_{,i}\over U}({\cal E}_0,\theta_0)
&=& \displaystyle\cases{
-{2\theta_{1,i}\over u_0(\pi-2\theta_1)}
(\lambda-\bar\lambda_\theta-n\lambda_\theta)^{-2}
+O\Bigl((\lambda-\bar\lambda_\theta-n\lambda_\theta)^0\Bigr) \,, \cr 
-{2\theta_{1,i}\over u_0(\pi-2\theta_1)}
(\lambda-\bar\lambda_\theta-(n+/2)\lambda_\theta)^{-2}
+O\Bigl((\lambda-\bar\lambda_\theta-(n+1/2)\lambda_\theta)^0\Bigr) \,.} 
\end{eqnarray}

We define regularization functions as 
\begin{eqnarray}
V^\dagger_i({\cal E}_0,\lambda-\bar\lambda_{r0}) 
&=& i{2\pi\over \lambda_r(r_2-r_1)}
\Biggl[{r_{1,i}\over v_1}
\biggl({1\over e^{i2\pi(\lambda-\bar\lambda_{r0})/\lambda_r}-1} 
-{1\over e^{-i2\pi(\lambda-\bar\lambda_{r0})/\lambda_r}-1}\biggr) 
\nonumber \\ && \qquad \qquad \qquad 
+{r_{2,i}\over v_2}
\biggl({1\over e^{i2\pi(\lambda-\bar\lambda_{r0})/\lambda_r}+1} 
-{1\over e^{-i2\pi(\lambda-\bar\lambda_{r0})/\lambda_r}+1}\biggr)\Biggr]
\,, \label{eq:d_r1} \\ 
U^\dagger_i({\cal E}_0,\lambda-\bar\lambda_{\theta 0}) 
&=& i{2\pi\over \lambda_\theta(r_2-r_1)}
{\theta_{1,i}\over u_0}\Biggl[
\biggl({1\over e^{i2\pi(\lambda-\bar\lambda_{\theta 0})/\lambda_\theta}-1} 
-{1\over e^{-i2\pi(\lambda-\bar\lambda_{\theta 0})/\lambda_\theta}-1}\biggr) 
\nonumber \\ && \qquad \qquad \qquad \qquad 
-\biggl({1\over e^{i2\pi(\lambda-\bar\lambda_{\theta 0})/\lambda_\theta}+1} 
-{1\over e^{-i2\pi(\lambda-\bar\lambda_{\theta 0})/\lambda_\theta}+1}
\biggr)\Biggr]
\,. \label{eq:d_the1} 
\end{eqnarray}
One can see that $dV^\dagger_i/d\lambda$ and $dU^\dagger/d\lambda$ 
have the same singular structures as $V_{,i}/2V$ and $U_{,i}/2U$, 
thus, $V_{,i}/2V-dV^\dagger_i/d\lambda$ 
and $U_{,i}/2U-dU^\dagger/d\lambda$ are regular and periodic 
with the period $\lambda_r$ and $\lambda_\theta$. 
One can expand these differences with discrete Fourier series 
$e^{i2\pi(\lambda-\bar\lambda_{r0})/\lambda_r}$ 
and $e^{i2\pi(\lambda-\bar\lambda_{\theta 0})/\lambda_\theta}$, 
and one can integrate in a regular manner. 
We formally write these integrations as 
\begin{eqnarray}
\Bigl[\bar V_i-V^\dagger_i\Bigr]({\cal E}_0,\lambda-\bar\lambda_{r 0}) 
&=& (\lambda-\bar\lambda_{r 0}) \dot V_i +\sum_n V^{(n)}_i({\cal E}_0)
e^{i2\pi n{\lambda-\bar\lambda_{r 0}\over \lambda_r}} 
\,, \label{eq:d_r2} \\ 
\Bigl[\bar U_i-U^\dagger_i\Bigr]({\cal E}_0,\lambda-\bar\lambda_{\theta 0}) 
&=& (\lambda-\bar\lambda_{\theta 0}) \dot U_i +\sum_n U^{(n)}_i({\cal E}_0)
e^{i2\pi n{\lambda-\bar\lambda_{\theta 0}\over \lambda_\theta}} 
\,, \label{eq:d_the2} 
\end{eqnarray}
where the coefficients of linearly growing terms are 
\begin{eqnarray}
\dot V_i({\cal E}_0) 
\,=\, {1\over\lambda_r}\int^{\lambda_r}_0 d\lambda 
\biggl({V_{,i}\over 2V}-{dV^\dagger_i\over d\lambda}\biggr) \,, \quad 
\dot U_i({\cal E}_0) \,=\, 
{1\over\lambda_\theta}\int^{\lambda_\theta}_0
d\lambda \biggl({U_{,i}\over 2U}-{dU^\dagger_i\over d\lambda}\biggr) 
\,. 
\end{eqnarray}
We note that there is an ambiguity in adding integral constants, 
which the finial result does not depend on. 

Using (\ref{eq:d_r1}), (\ref{eq:d_the1}), (\ref{eq:d_r2}) 
and (\ref{eq:d_the2}), 
the first terms of (\ref{eq:d_r0}) and (\ref{eq:d_the0}) 
can be separeted as 
\begin{eqnarray} 
{dr_0\over d\lambda}
\biggl([\bar V_i-V^\dagger_i]\delta {\cal E}^i
+V^\dagger_i\delta {\cal E}^i\biggr) 
\,, \quad 
{d\theta_0\over d\lambda}
\biggl([\bar U_i-U^\dagger_i]\delta {\cal E}^i 
+U^\dagger_i\delta {\cal E}^i\biggr) \,. \nonumber 
\end{eqnarray} 
Now we can see that we successfully regularize 
the first terms of (\ref{eq:d_r0}) and (\ref{eq:d_the0}) 
since the divergent behavior of $V^\dagger_i$ and $U^\dagger_i$ 
and the regular behavior of $dr_0/d\lambda$ and $d\theta_0/d\lambda$ 
cancel each other at the singularities of (\ref{eq:d_r_the}) as 
\begin{eqnarray}
{dr_0\over d\lambda}V^\dagger_i
&=& \displaystyle\cases{
r_{1,i}+O\Bigl((\lambda-\bar\lambda_{r0}-n\lambda_r)^2\Bigr) \,, \cr 
r_{2,i}+O\Bigl((\lambda-\bar\lambda_{r0}-(n+1/2)\lambda_r)^2\Bigr) \,,} 
\\ 
{d\theta_0\over d\lambda}U^\dagger_i
&=& \displaystyle\cases{
\theta_{1,i}
+O\Bigl((\lambda-\bar\lambda_{\theta 0}-n\lambda_\theta)^2\Bigr) \,, \cr 
-\theta_{1,i}
+O\Bigl((\lambda-\bar\lambda_{\theta 0}-(n+1/2)\lambda_\theta)^2\Bigr) \,.} 
\end{eqnarray}

We can further rewrite 
the first terms of (\ref{eq:d_r0}) and (\ref{eq:d_the0}) 
using $R_{,i}({\cal E}_0,\lambda-\bar\lambda_{r 0})$ 
and $\Theta_{,i}({\cal E}_0,\lambda-\bar\lambda_{\theta 0})$. 
By taking the ${\cal E}$ derivative of the geodesic equation, 
we have 
\begin{eqnarray}
{d\over d\lambda}\left({R_{,i}\over{dr_0\over d\lambda}}\right) 
\,=\, {1\over 2}{V_{,i}\over V}({\cal E}_0,r_0) \,, \quad 
{d\over d\lambda}\left({\Theta_{,i}\over{d\theta_0\over d\lambda}}\right) 
\,=\, {1\over 2}{U_{,i}\over U}({\cal E}_0,\theta_0) \,.
\end{eqnarray}
These equations have singularities as (\ref{eq:d_r_the}), 
and must be integrated 
such that $R_{,i}$ and $\Theta_{,i}$ are smooth at the singular points. 
Using (\ref{eq:d_r1}), (\ref{eq:d_the1}), 
(\ref{eq:d_r2}) and (\ref{eq:d_the2}), we have 
\begin{eqnarray} 
R_{,i} ({\cal E}_0,\lambda-\bar\lambda_{r 0}) 
\,=\, {dr_0\over d\lambda}
\biggl([\bar V_i-V^\dagger_i]+V^\dagger_i+c^{(v)}_i\biggr) \,, \quad 
\Theta_{,i}({\cal E}_0,\lambda-\bar\lambda_{\theta 0}) 
\,=\, {d\theta_0\over d\lambda}
\biggl([\bar U_i-U^\dagger_i]+U^\dagger_i+c^{(u)}_i\biggr) \,, 
\label{eq:r_the_diff}
\end{eqnarray} 
where $c^{(v)}_i$ and $c^{(u)}_i$ are finite integral constants. 
Using the ambiguity of integral constants 
in evaluating $[\bar V_i-V^\dagger_i]$ and $[\bar U_i-U^\dagger_i]$, 
we set $c^{(v)}_i=c^{(u)}_i=0$. 
and we have the first terms of (\ref{eq:d_r0}) and (\ref{eq:d_the0}) as 
\begin{eqnarray} 
R_{,i}({\cal E}_0,\lambda-\bar\lambda_{r 0})\delta {\cal E}^i \,, \quad 
\Theta_{,i}({\cal E}_0,\lambda-\bar\lambda_{\theta 0})\delta {\cal E}^i \,. 
\label{eq:d_r_the1} 
\end{eqnarray} 

Since $r$- and $\theta$-motion of a geodesic is periodic, 
we can put 
\begin{eqnarray}
R({\cal E},\lambda-\bar\lambda_r) \,=\, \sum_n R^{(n)}({\cal E})
e^{i2\pi n{\lambda-\bar\lambda_r\over\lambda_r({\cal E})}} 
\,, \quad 
\Theta({\cal E},\lambda-\bar\lambda_r) \,=\, \sum_n \Theta^{(n)}({\cal E})
e^{i2\pi n{\lambda-\bar\lambda_\theta\over\lambda_\theta({\cal E})}} 
\,. 
\end{eqnarray}
The ${\cal E}$-derivative of $R$ and $\Theta$ becomes 
\begin{eqnarray}
R_{,i} &=& \sum_n \biggl(
-i2\pi n{\lambda-\bar\lambda_r\over\lambda_r}
{\lambda_{r,i}\over\lambda_r}R^{(n)}+R^{(n)}_{,i}
\biggr)e^{i2\pi n{\lambda-\bar\lambda_r\over\lambda_r}} 
\,=\, -(\lambda-\bar\lambda_r){dR\over d\lambda}
{\lambda_{r,i}\over\lambda_r}
+\sum_n R^{(n)}_{,i} e^{i2\pi n{\lambda-\bar\lambda_r\over\lambda_r}} 
\,, \label{eq:r_i} \\ 
\Theta_{,i} &=& \sum_n \biggl(
-i2\pi n{\lambda-\bar\lambda_\theta\over\lambda_\theta}
{\lambda_{\theta,i}\over\lambda_\theta}\Theta^{(n)}+\Theta^{(n)}_{,i}
\biggr)e^{i2\pi n{\lambda-\bar\lambda_\theta\over\lambda_\theta}} 
\,=\, -(\lambda-\bar\lambda_\theta){d\Theta\over d\lambda} 
{\lambda_{\theta,i}\over\lambda_\theta}
+\sum_n \Theta^{(n)}_{,i} 
e^{i2\pi n{\lambda-\bar\lambda_\theta\over\lambda_\theta}} 
\,. \label{eq:the_i}
\end{eqnarray}
One can see that $R_{,i}$ and $\Theta_{,}$ are 
dominated by oscillating parts 
whose amplitude grows linearly in $\lambda$. 
Comparing (\ref{eq:d_r2}), (\ref{eq:d_the2}) and (\ref{eq:r_the_diff}), 
we find 
\begin{eqnarray}
\dot V_i({\cal E}_0) 
\,=\, -{\lambda_{r,i}\over\lambda_r}({\cal E}_0) 
\,, \quad 
\dot U_i({\cal E}_0) 
\,=\, -{\lambda_{\theta,i}\over\lambda_\theta}({\cal E}_0) 
\,. 
\end{eqnarray}

We next discuss the second terms of (\ref{eq:d_r0}) and (\ref{eq:d_the0}). 
Using (\ref{eq:d_r1}), (\ref{eq:d_the1}) 
(\ref{eq:d_r2}) and (\ref{eq:d_the2}), 
we separete the terms as 
\begin{eqnarray} 
-{dr_0\over d\lambda}
\biggl(\int^\lambda_0 d\lambda [\bar V_i-V^\dagger_i]
{d\over d\lambda}\delta {\cal E}^i
+\int^\lambda_0 d\lambda V^\dagger_i
{d\over d\lambda}\delta {\cal E}^i\biggr) 
\,, \label{eq:d_r3} \\ 
-{d\theta_0\over d\lambda}
\biggl(\int^\lambda_0 d\lambda [\bar U_i-U^\dagger_i]
{d\over d\lambda}\delta {\cal E}^i 
+\int^\lambda_0 d\lambda_0 U^\dagger_i
{d\over d\lambda}\delta {\cal E}^i\biggr) 
\,. \label{eq:d_the3} 
\end{eqnarray}

The integrands of the first terms in the brackets are regular and periodic, 
and one can evaluate the integration in an usual manner as 
\begin{eqnarray} 
\int^\lambda_0 d\lambda [\bar V_i-V^\dagger_i]
{d\over d\lambda}\delta {\cal E}^i
&=& {(\lambda-\bar\lambda_{r 0})^2\over 2}\ddot V_e 
+\sum_{m,n}\biggl((\lambda-\bar\lambda_{r 0})\dot V_e^{(m,n)}
+V_e^{(m,n)}\biggr)
e^{i2\pi\bigl(m{\lambda-\bar\lambda_{r 0}\over\lambda_r}
+n{\lambda-\bar\lambda_{\theta 0}\over\lambda_\theta}\bigr)} 
\,, \label{eq:d_r4} \\ 
\int^\lambda_0 d\lambda [\bar U_i-U^\dagger_i]
{d\over d\lambda}\delta {\cal E}^i 
&=& {(\lambda-\bar\lambda_{\theta 0})^2\over 2}\ddot U_e 
+\sum_{m,n}\biggr((\lambda-\bar\lambda_{\theta 0})\dot U_e^{(m,n)}
+U_e^{(m,n)}\biggr)
e^{i2\pi\bigl(m{\lambda-\bar\lambda_{r 0}\over\lambda_r}
+n{\lambda-\bar\lambda_{\theta 0}\over\lambda_\theta}\bigr)} 
\,. \label{eq:d_the4} 
\end{eqnarray}

The second terms in the brackets could lead to logarithmic divergence 
at the singular points, $r_0=r_1,r_2$ and $\theta_0=\theta_1,\pi-\theta_1$. 
If we have logarithmic divergence, 
we have no way to have a smooth evolution of the orbit at the singularities. 
Thus, one must constraint on the self-force as 
\begin{eqnarray}
0 &=& r_{1,i}{d\over d\lambda}\delta{\cal E}^i
(\lambda=\bar\lambda_{r 0}+n\lambda_r) 
\,=\, r_{2,i}{d\over d\lambda}\delta{\cal E}^i
(\lambda=\bar\lambda_{r 0}+(n+1/2)\lambda_r) 
\,, \label{eq:r_reg} \\ 
0 &=& \theta_{1,i}{d\over d\lambda}\delta{\cal E}^i
(\lambda=\bar\lambda_{\theta 0}+n\lambda_\theta) 
\,=\, \theta_{1,i}{d\over d\lambda}\delta{\cal E}^i
(\lambda=\bar\lambda_{\theta 0}+(n+1/2)\lambda_\theta) 
\,. \label{eq:the_reg} 
\end{eqnarray}
Using (\ref{eq:g_per}), we can re-expand as 
\begin{eqnarray} 
r_{1,i}{d\over d\lambda}\delta{\cal E}^i
&=& \sum_{m \not =0,n} \dot {\cal E}_{r1}^{(m,n)}
(e^{i2m\pi{\lambda-\bar\lambda_{r 0}\over\lambda_r}}-1)
e^{i2n\pi{\lambda-\bar\lambda_{\theta 0}\over\lambda_\theta}} 
\,, \\ 
r_{2,i}{d\over d\lambda}\delta{\cal E}^i
&=& \sum_{m \not =0,n} \dot {\cal E}_{r2}^{(m,n)}
((-1)^me^{i2m\pi{\lambda-\bar\lambda_{r 0}\over\lambda_r}}-1)
e^{i2n\pi{\lambda-\bar\lambda_{\theta 0}\over\lambda_\theta}} 
\,, \\ 
\theta_{1,i}{d\over d\lambda}\delta{\cal E}^i
&=& \sum_{m,n \not =0} \dot {\cal E}_{\theta 1}^{(m,n)} 
e^{i2m\pi{\lambda-\bar\lambda_{r 0}\over\lambda_r}}
(e^{i2n\pi{\lambda-\bar\lambda_{\theta 0}\over\lambda_\theta}}-1) 
\,, \\ 
-\theta_{1,i}{d\over d\lambda}\delta{\cal E}^i
&=& \sum_{m \not =0,n} \dot {\cal E}_{\theta 2}^{(m,n)} 
e^{i2m\pi{\lambda-\bar\lambda_{r 0}\over\lambda_r}} 
((-1)^me^{i2n\pi{\lambda-\bar\lambda_{\theta 0}\over\lambda_\theta}}-1)
\,, 
\end{eqnarray} 
where $\dot {\cal E}_{r1}^{(m,n)}=r_{1,i}\dot {\cal E}^{i(m,n)}$, 
$\dot {\cal E}_{r2}^{(m,n)}=(-1)^mr_{2,i}\dot {\cal E}^{i(m,n)}$, 
$\dot {\cal E}_{\theta 1}^{(m,n)}=\theta_{1,i}\dot {\cal E}^{i(m,n)}$ and 
$\dot {\cal E}_{\theta 2}^{(m,n)}=-(-1)^n\theta_{1,i}\dot {\cal E}^{i(m,n)}$. 
Using these expansions, we can integrate the second term 
without logarithmic divergence as 
\begin{eqnarray} 
\int^\lambda_0 d\lambda V^\dagger_i{d\over d\lambda}\delta {\cal E}^i
&=& (\lambda-\bar\lambda_{r 0})\dot V_e^\dagger
+\sum_{m,n}V_e^{\dagger(m,n)}
e^{i2\pi\bigl(m{\lambda-\bar\lambda_{r 0}\over\lambda_r}
+n{\lambda-\bar\lambda_{\theta 0}\over\lambda_\theta}\bigr)} 
\,, \label{eq:d_r5} \\ 
\int^\lambda_0 d\lambda U^\dagger_i{d\over d\lambda}\delta {\cal E}^i
&=& (\lambda-\bar\lambda_{\theta 0})\dot U_e^\dagger
+\sum_{m,n}U_e^{\dagger(m,n)}
e^{i2\pi\bigl(m{\lambda-\bar\lambda_{r 0}\over\lambda_r}
+n{\lambda-\bar\lambda_{\theta 0}\over\lambda_\theta}\bigr)} 
\,. \label{eq:d_the5}
\end{eqnarray}
It is notable that we have only linearly growing terms, 
thus, (\ref{eq:d_r4}) and (\ref{eq:d_the4}) dominate 
over (\ref{eq:d_r5}) and (\ref{eq:d_the5}) 
at $\lambda\sim O(\mu^\alpha),\,0>\alpha>-1/2$. 

Using (\ref{eq:r_the_diff}), one can rewrite 
the second terms of (\ref{eq:d_r0}) and (\ref{eq:d_the0}) as 
\begin{eqnarray} 
-{dr_0\over d\lambda}
\int^\lambda_0 d\lambda {R_{,i}\over {dr_0\over d\lambda}}
{d\over d\lambda}\delta {\cal E}^i
\,, \quad 
-{d\theta_0\over d\lambda}
\int^\lambda_0 d\lambda {\Theta_{,i}\over {d\theta_0\over d\lambda}}
{d\over d\lambda}\delta {\cal E}^i 
\,. \label{eq:d_r_the2} 
\end{eqnarray}
By the constraints on the self-force 
(\ref{eq:r_reg}) and (\ref{eq:the_reg}), 
the integrands of (\ref{eq:d_r_the2}) have no singularity, 
and we can formally write as 
\begin{eqnarray} 
{R_{,i}\over {dr_0\over d\lambda}}{d\over d\lambda}\delta {\cal E}^i
&=& \sum_{m,n}
\biggl((\lambda-\bar\lambda_{r0})\dot R_e^{(m,n)}
+R_e^{(m,n)}\biggr) 
e^{i2\pi\bigl(m{\lambda-\bar\lambda_{r0}\over\lambda_r}
+n{\lambda-\bar\lambda_{\theta 0}\over\lambda_\theta}\bigr)} 
\,, \label{eq:d_r6} \\ 
{\Theta_{,i}\over {d\theta_0\over d\lambda}}
{d\over d\lambda}\delta {\cal E}^i
&=& \sum_{m,n}
\biggl((\lambda-\bar\lambda_{\theta 0})\dot \Theta_e^{(m,n)}
+\Theta_e^{(m,n)}\biggr) 
e^{i2\pi\bigl(m{\lambda-\bar\lambda_{r0}\over\lambda_r}
+n{\lambda-\bar\lambda_{\theta 0}\over\lambda_\theta}\bigr)} 
\,, \label{eq:d_the6} 
\end{eqnarray}
where the coefficients of linearly growing terms become 
\begin{eqnarray}
\dot R_e^{(m,n)} \,=\, -{\lambda_{r,i}\over\lambda_r}
\dot {\cal E}^{R-ret.\,(m,n)\,i}
\,, \quad 
\dot \Theta_e^{(m,n)} \,=\, -{\lambda_{\theta,i}\over\lambda_\theta}
\dot {\cal E}^{R-ret.\,(m,n)\,i}
\,. \label{eq:d_r_the3} 
\end{eqnarray}

We show how to integrate (\ref{eq:d_r0}) and (\ref{eq:d_the0}) 
such that $\delta r$ and $\delta\theta$ evolve smoothly, 
and we find $c_v^{(n)}=c_u^{(n)}=0$ along this integration procedure. 
The perturbative evolution of $\delta r$ and $\delta\theta$ by the self-force 
is now interpreted as the evolution of the orbital `constants', 
$\delta {\cal E}^i$, $\delta\bar\lambda_r$ and $\delta\bar\lambda_\theta$ as 
\begin{eqnarray} 
\delta r 
&=& R_{,i}\delta {\cal E}^i+R_{,\bar\lambda_r}\delta\bar\lambda_r
\,, \label{eq:d_r} \\ 
\delta\bar\lambda_r 
&=& {(\lambda-\bar\lambda_{r0})^2\over 2} \ddot\lambda_r 
+\sum_{m,n}\biggl((\lambda-\bar\lambda_{r0})\dot\lambda_r^{(m,n)}
+\lambda_r^{(m,n)}\biggr) 
e^{i2\pi\bigl(m{\lambda-\bar\lambda_{r0}\over\lambda_r}
+n{\lambda-\bar\lambda_{\theta 0}\over\lambda_\theta}\bigr)} 
\,, \label{eq:l_r} \\ 
\delta\theta 
&=& \Theta_{,i}\delta {\cal E}^i
+\Theta_{,\bar\lambda_\theta}\delta\bar\lambda_\theta
\,, \label{eq:d_the} \\ 
\delta\bar\lambda_\theta 
&=& {(\lambda-\bar\lambda_{\theta 0})^2\over 2} \ddot\lambda_\theta 
+\sum_{m,n}\biggl((\lambda-\bar\lambda_{\theta 0})\dot\lambda_\theta^{(m,n)}
+\lambda_\theta^{(m,n)}\biggr) 
e^{i2\pi\bigl(m{\lambda-\bar\lambda_{r 0}\over\lambda_r}
+n{\lambda-\bar\lambda_{\theta 0}\over\lambda_\theta}\bigr)} 
\,. \label{eq:l_the} 
\end{eqnarray}
From (\ref{eq:d_r_the3}), we have 
\begin{eqnarray}
\ddot\lambda_r &=& 
\,=\, -{\lambda_{r,i}\over\lambda_r}\dot{\cal E}^{R-ret.\,(0,0)\,i}
\,, \quad 
\ddot\lambda_\theta 
\,=\, -{\lambda_{\theta,i}\over\lambda_\theta}\dot{\cal E}^{R-ret.\,(0,0)\,i}
\,, \\ 
\dot\lambda_r^{(m,n)} 
&=& {i\over 2\pi\bigl({m\over\lambda_r}+{n\over\lambda_\theta}\bigr)}
{\lambda_{r,i}\over\lambda_r}\dot{\cal E}^{R-ret.\,(m,n)\,i} 
\,, \quad 
\dot\lambda_\theta^{(m,n)} 
\,=\, {i\over 2\pi\bigl({m\over\lambda_r}+{n\over\lambda_\theta}\bigr)}
{\lambda_{\theta,i}\over\lambda_\theta}\dot{\cal E}^{R-ret.\,(m,n)\,i} 
\,, \\ 
\dot\lambda_r^{(0,0)} 
&=& R_e^{(0,0)} \,=\, \dot V_e^{(0,0)}+\dot V_e^\dagger 
\,, \quad 
\dot\lambda_\theta^{(0,0)} 
\,=\, \Theta_e^{(0,0)} \,=\, \dot U_e^{(0,0)}+\dot U_e^\dagger 
\,, \\ 
\lambda_r^{(m,n)} 
&=& {-i\over 2\pi\bigl({m\over\lambda_r}+{n\over\lambda_\theta}\bigr)}
R_e^{(m,n)}
\,=\, V_e^{(m,n)}+V_e^{\dagger(m,n)} 
\,, \quad 
\lambda_\theta^{(m,n)} 
\,=\, {-i\over 2\pi\bigl({m\over\lambda_r}+{n\over\lambda_\theta}\bigr)}
\Theta_e^{(m,n)}
\,=\, U_e^{(m,n)}+U_e^{\dagger(m,n)} 
\,, 
\end{eqnarray}
where the second line and the fourth line are 
evaluated for $(m,n)\not =(0,0)$, 
and $\lambda_r^{(0,0)}$, $\lambda_\theta^{(0,0)}$ are determined 
such that $\delta\bar\lambda_0=\delta\bar\lambda_\theta=0$ at $\lambda=0$.

We can see that the linear perturbation holds 
for $\lambda\sim O(\mu^\alpha),\,0\geq\alpha>-1/2$. 
The orbital evolution in this regime has two parts; 
the one contributed by $\delta {\cal E}$, 
the other by $\delta\bar\lambda_r$ and $\delta\bar\lambda_\theta$. 
Since both parts grow quadratically by $\lambda$, 
it is necessary to consider 
the evolution of $\bar\lambda_r$ and $\bar\lambda_\theta$ 
for a correct orbital prediction.

\subsection{$t$-motion and $\phi$-motion} \label{sec:t-p} 

We write (\ref{eq:g_t}) and (\ref{eq:g_p}) as 
\begin{eqnarray}
{dt \over d\lambda} \,=\, X({\cal E},r,\theta) \,, \quad 
{d\phi \over d\lambda} \,=\, Y({\cal E},r,\theta) \,. 
\label{eq:t_p} 
\end{eqnarray}
The leading deviation from the initial geodesic satisfies 
\begin{eqnarray}
{d\delta t \over d\lambda} 
&=& X_{,i}({\cal E}_0,r_0,\theta_0)\delta {\cal E}^i
+X_{,r}({\cal E}_0,r_0,\theta_0)\delta r
+X_{,\theta}({\cal E}_0,r_0,\theta_0)\delta \theta 
\nonumber \\ 
&=& (X_{,i}+X_{,r}R_{,i}+X_{,\theta}\Theta_{,i})\delta {\cal E}^i
+X_{,r}R_{,\bar\lambda_r}\delta\bar\lambda_r
+X_{,\theta}\Theta_{,\bar\lambda_\theta}\delta\bar\lambda_\theta
\,, \label{eq:d_t0} \\ 
{d\delta\phi \over d\lambda} 
&=& Y_{,i}({\cal E}_0,r_0,\theta_0)\delta {\cal E}^i
+Y_{,r}({\cal E}_0,r_0,\theta_0)\delta r
+Y_{,\theta}({\cal E}_0,r_0,\theta_0)\delta\theta 
\nonumber \\ 
&=& (Y_{,i}+Y_{,r}R_{,i}+Y_{,\theta}\Theta_{,i})\delta {\cal E}^i
+Y_{,r}R_{,\bar\lambda_r}\delta\bar\lambda_r
+Y_{,\theta}\Theta_{,\bar\lambda_\theta}\delta\bar\lambda_\theta
\,, \label{eq:d_p0} 
\end{eqnarray}
where we use (\ref{eq:d_r}) and (\ref{eq:d_the}). 
Contrary to (\ref{eq:d_r_the}), 
(\ref{eq:d_t0}) and (\ref{eq:d_p0}) are regular, 
and we can integrate by parts as 
\begin{eqnarray}
\delta t &=& \bar X_i\delta {\cal E}^i
+\bar X_r\delta\bar\lambda_r+\bar X_\theta\delta\bar\lambda_\theta
-\int^\lambda_0 d\lambda \biggl(
\bar X_i{d\over d\lambda}\delta{\cal E}^i
+\bar X_r{d\over d\lambda}\delta\bar\lambda_r
+\bar X_\theta{d\over d\lambda}\delta\bar\lambda_\theta
\biggr)
\,, \label{eq:d_t1} \\ 
{d\over d\lambda}\bar X_i
&=& X_{,i}+X_{,r}R_{,i}+X_{,\theta}\Theta_{,i} \,, \quad 
{d\over d\lambda}\bar X_r 
\,=\, X_{,r}R_{,\bar\lambda_r} \,, \quad 
{d\over d\lambda}\bar X_\theta 
\,=\, X_{,\theta}\Theta_{,\bar\lambda_\theta} 
\,, \label{eq:d_t2} \\ 
\delta\phi &=& \bar Y_i\delta {\cal E}^i
+\bar Y_r\delta\bar\lambda_r+\bar Y_\theta\delta\bar\lambda_\theta
-\int^\lambda_0 d\lambda 
\biggl(\bar Y_i{d\over d\lambda}\delta{\cal E}^i
+\bar Y_r{d\over d\lambda}\delta\bar\lambda_r
+\bar Y_\theta{d\over d\lambda}\delta\bar\lambda_\theta\biggr)
\,, \label{eq:d_p1} \\ 
{d\over d\lambda}\bar Y_i
&=& Y_{,i}+Y_{,r}R_{,i}+Y_{,\theta}\Theta_{,i} \,, \quad 
{d\over d\lambda}\bar Y_r 
\,=\, Y_{,r}R_{,\bar\lambda_r} \,, \quad 
{d\over d\lambda}\bar Y_\theta 
\,=\, Y_{,\theta}\Theta_{,\bar\lambda_\theta} 
\,. \label{eq:d_p2} 
\end{eqnarray}

We note that 
$T_{,i}({\cal E}_0,\bar\lambda_{r0},\bar\lambda_{\theta 0};\lambda)$ 
and $\Phi_{,i}({\cal E}_0,\bar\lambda_{r0},\bar\lambda_{\theta 0};\lambda)$ 
satisfy 
\begin{eqnarray}
{d\over d\lambda}T_{,i}
\,=\, X_{,i}+X_{,r}R_{,i}+X_{,\theta}\Theta_{,i} \,, \quad 
{d\over d\lambda}\Phi_{,i}
\,=\, X_{,i}+X_{,r}R_{,i}+X_{,\theta}\Theta_{,i} \,, 
\end{eqnarray}
thus, from (\ref{eq:d_t2}) and (\ref{eq:d_p2}), we have 
\begin{eqnarray}
\bar X_i \,=\, T_{,i}({\cal E}_0,\bar\lambda_{r0},\bar\lambda_{\theta 0})
+c^{(x)} \,, \quad 
\bar Y_i \,=\, \Phi_{,i}({\cal E}_0,\bar\lambda_{r0},\bar\lambda_{\theta 0})
+c^{(y)} \,, 
\label{eq:xy_i} 
\end{eqnarray} 
where $c^{(x)}$ and $c^{(y)}$ are integral constants. 
Because the orbital evolution does not depend on 
$c^{(x)}$ and $c^{(y)}$ in the end, 
we set them zero in this subsection. 
In the same manner, we have 
\begin{eqnarray}
\bar X_r &=& T_{,\bar\lambda_r} \,, \quad 
\bar X_\theta \,=\, T_{,\bar\lambda_\theta} \,, \quad 
\bar Y_r \,=\, \Phi_{,\bar\lambda_r} \,, \quad 
\bar Y_\theta \,=\, \Phi_{,\bar\lambda_\theta} \,. 
\label{eq:xy_r_the} 
\end{eqnarray} 

We define $t$- and $\phi$-motion of a geodesic as 
\begin{eqnarray}
T({\cal E},\bar\lambda_r,\bar\lambda_\theta;\lambda) 
&=& \lambda \dot T({\cal E}) +\sum_{m,n} T^{(m,n)}({\cal E})
e^{i2\pi \bigl(m{\lambda-\bar\lambda_r\over\lambda_r({\cal E})}
+n{\lambda-\bar\lambda_\theta\over\lambda_\theta({\cal E})}\bigr)} 
\,, \\ 
\Phi({\cal E},\bar\lambda_r,\bar\lambda_\theta;\lambda) 
&=& \lambda \dot \Phi({\cal E}) +\sum_{m,n} \Phi^{(m,n)}({\cal E})
e^{i2\pi \bigl(m{\lambda-\bar\lambda_r\over\lambda_r({\cal E})}
+n{\lambda-\bar\lambda_\theta\over\lambda_\theta({\cal E})}\bigr)} 
\,, 
\end{eqnarray}
where the linearly growing terms appear 
since we integrate $X({\cal E},R,\Theta)$ and $Y({\cal E},R,\Theta)$ 
which can be expanded by discrete Fourier-series 
$e^{i2\pi m(\lambda-\bar\lambda_r)/\lambda_r
+i2\pi n(\lambda-\bar\lambda_\theta)/\lambda_\theta}$. 
Using (\ref{eq:xy_i}) and (\ref{eq:xy_r_the}), 
we have 
\begin{eqnarray}
\bar X_i &=& \sum_{m,n} 
\bigl(\lambda \dot X_i^{(m,n)} +X_i^{(m,n)}\bigr)
e^{i2\pi \bigl(m{\lambda-\bar\lambda_{r 0}\over\lambda_r}
+n{\lambda-\bar\lambda_{\theta 0}\over\lambda_\theta}\bigr)} 
\,, \quad 
\bar Y_i \,=\, \sum_{m,n} 
\bigl(\lambda \dot Y_i^{(m,n)} +Y_i^{(m,n)}\bigr)
e^{i2\pi \bigl(m{\lambda-\bar\lambda_{r 0}\over\lambda_r}
+n{\lambda-\bar\lambda_{\theta 0}\over\lambda_\theta}\bigr)} 
\,, \\ 
\bar X_r &=& \sum_{m,n} X_r^{(m,n)} 
e^{i2\pi \bigl(m{\lambda-\bar\lambda_{r 0}\over\lambda_r}
+n{\lambda-\bar\lambda_{\theta 0}\over\lambda_\theta}\bigr)} 
\,, \quad 
\bar Y_r \,=\, \sum_{m,n} Y_r^{(m,n)} 
e^{i2\pi \bigl(m{\lambda-\bar\lambda_{r 0}\over\lambda_r}
+n{\lambda-\bar\lambda_{\theta 0}\over\lambda_\theta}\bigr)} 
\,, \\ 
\bar X_\theta &=& \sum_{m,n} X_\theta^{(m,n)} 
e^{i2\pi \bigl(m{\lambda-\bar\lambda_{r 0}\over\lambda_r}
+n{\lambda-\bar\lambda_{\theta 0}\over\lambda_\theta}\bigr)} 
\,, \quad  
\bar Y_\theta \,=\, \sum_{m,n} Y_\theta^{(m,n)} 
e^{i2\pi \bigl(m{\lambda-\bar\lambda_{r 0}\over\lambda_r}
+n{\lambda-\bar\lambda_{\theta 0}\over\lambda_\theta}\bigr)} 
\,, 
\end{eqnarray}
where the coefficients become 
\begin{eqnarray}
\dot X_i^{(m,n)} &=& \delta_{m=n=0} \dot T_{,i}
-i2\pi \biggl(m{\lambda_{r,i}\over \lambda_r^2}
+n{\lambda_{\theta,i}\over \lambda_\theta^2}\biggr) T^{(m,n)} 
\,, \quad 
X_i^{(m,n)} \,=\, T^{(m,n)}_{,i}
+i2\pi \biggl(m{\bar\lambda_r\lambda_{r,i}\over \lambda_r^2}
+n{\bar\lambda_\theta\lambda_{\theta,i}\over \lambda_\theta^2}\biggr)
T^{(m,n)} 
\,, \\ 
\dot Y_i^{(m,n)} &=& \delta_{m=n=0} \dot \Phi_{,i}
-i2\pi \biggl(m{\lambda_{r,i}\over \lambda_r^2}
+n{\lambda_{\theta,i}\over \lambda_\theta^2}\biggr) \Phi^{(m,n)} 
\,, \quad 
Y_i^{(m,n)} \,=\, \Phi^{(m,n)}_{,i} 
+i2\pi \biggl(m{\bar\lambda_r\lambda_{r,i}\over \lambda_r^2}
+n{\bar\lambda_\theta\lambda_{\theta,i}\over \lambda_\theta^2}\biggr)
\Phi^{(m,n)} 
\,, \\ 
X_r^{(m,n)} &=& -m{i2\pi \over\lambda_r} T^{(m,n)} 
\,, \quad 
X_\theta^{(m,n)} \,=\, -n{i2\pi \over\lambda_\theta} T^{(m,n)} 
\,, \quad 
Y_r^{(m,n)} \,=\, -m{i2\pi \over\lambda_r} \Phi^{(m,n)} 
\,, \quad 
Y_\theta^{(m,n)} \,=\, -n{i2\pi \over\lambda_\theta} \Phi^{(m,n)} 
\,. 
\end{eqnarray}

We consider to evaluate 
the integration of (\ref{eq:d_t1}) and (\ref{eq:d_p1}). 
We first note 
\begin{eqnarray}
{d\over d\lambda}\delta\bar\lambda_r &=& 
{R_{,i}\over{dr_0\over d\lambda}}{d\over d\lambda}\delta{\cal E}^i 
\,, \quad 
{d\over d\lambda}\delta\bar\lambda_\theta \,=\, 
{\Theta_{,i}\over{d\theta_0\over d\lambda}}{d\over d\lambda}\delta{\cal E}^i 
\,, 
\end{eqnarray}
and the expansion of RHSs are given 
in (\ref{eq:d_r5}), (\ref{eq:d_the5}) and (\ref{eq:d_r_the3}). 
Using these results, 
the integrands of (\ref{eq:d_t1}) and (\ref{eq:d_p1}) become 
\begin{eqnarray}
\sum_{m,n}(\lambda \dot X_e^{(m,n)}+X_e^{(m,n)})
e^{i2\pi \bigl(m{\lambda-\bar\lambda_{r 0}\over\lambda_r}
+n{\lambda-\bar\lambda_{\theta 0}\over\lambda_\theta}\bigr)} 
\,, \quad 
\sum_{m,n}(\lambda \dot Y_e^{(m,n)}+Y_e^{(m,n)})
e^{i2\pi \bigl(m{\lambda-\bar\lambda_{r 0}\over\lambda_r}
+n{\lambda-\bar\lambda_{\theta 0}\over\lambda_\theta}\bigr)} 
\,, 
\end{eqnarray} 
where the coefficients of the linearly growing and oscillating terms become 
\begin{eqnarray}
\dot X_e^{(m,n)} \,=\, \dot T_{,i}\dot{\cal E}^{R-ret.\,(m,n)\,i} 
\,, \quad 
\dot Y_e^{(m,n)} \,=\, \dot \Phi_{,i}\dot{\cal E}^{R-ret.\,(m,n)\,i} 
\,. 
\end{eqnarray}

In summary, we have $t$- and $\phi$-motion as 
\begin{eqnarray}
\delta t &=& T_{,i}\delta {\cal E}^i
+T_{,\bar\lambda_r}\delta\bar\lambda_r
+T_{,\bar\lambda_\theta}\delta\bar\lambda_\theta
+\delta \bar t 
\,, \label{eq:d_t} \\ 
\delta \bar t &=& {\lambda^2\over 2}\ddot t_0 
+\sum_{m,n}(\lambda \dot t_0^{(m,n)}+t_0^{(m,n)})
e^{i2\pi \bigl(m{\lambda-\bar\lambda_{r 0}\over\lambda_r}
+n{\lambda-\bar\lambda_{\theta 0}\over\lambda_\theta}\bigr)} 
\,, \\ 
\delta\phi &=& \Phi_{,i}\delta {\cal E}^i
+\Phi_{,\bar\lambda_r}\delta\bar\lambda_r
+\Phi_{,\bar\lambda_\theta}\delta\bar\lambda_\theta
+\delta \bar \phi 
\,, \label{eq:d_p} \\ 
\delta \bar \phi &=& {\lambda^2\over 2}\ddot \phi_0 
+\sum_{m,n}(\lambda \dot \phi_0^{(m,n)}+\phi_0^{(m,n)})
e^{i2\pi \bigl(m{\lambda-\bar\lambda_{r 0}\over\lambda_r}
+n{\lambda-\bar\lambda_{\theta 0}\over\lambda_\theta}\bigr)} 
\,, 
\end{eqnarray}
where the part of coefficients of $\delta \bar t$ and $\delta \bar\phi$ are 
\begin{eqnarray}
\ddot t_0 &=& -\dot T_{,i}\dot{\cal E}^{R-ret.\,(0,0)\,i} 
\,, \quad 
\ddot \phi_0 \,=\, -\dot \Phi_{,i}\dot{\cal E}^{R-ret.\,(0,0)\,i} 
\,, \\ 
\dot t_0^{(m,n)} &=& 
{i\over 2\pi \bigl({m\over\lambda_r}+{n\over\lambda_\theta}\bigr)} 
\dot T_{,i}\dot{\cal E}^{R-ret.\,(m,n)\,i} 
\,, \quad 
\dot \phi_0^{(m,n)} \,=\, 
{i\over 2\pi \bigl({m\over\lambda_r}+{n\over\lambda_\theta}\bigr)} 
\dot \Phi_{,i}\dot{\cal E}^{R-ret.\,(m,n)\,i} 
\,, \quad (m,n) \not = (0,0) \,. 
\end{eqnarray}

As in the case of $r$- and $\theta$-motion, 
the linear perturbation holds 
for $\lambda\sim O(\mu^\alpha),\,0\geq\alpha>-1/2$\footnote{
This time scale is called the dephasing time.\cite{dephase}}, 
and $\delta {\cal E}$, $\delta\bar\lambda_r$, $\delta\bar\lambda_\theta$, 
$\delta\bar t$ and $\delta\bar\phi$ 
contribute to the orbital evolution equally.

\subsection{Adiabatic Evolution} \label{sec:adi}

In Subsec.\ref{sec:r-the} and Subsec.\ref{sec:t-p}, 
we find the evolution of the orbit by the self-force 
in a perturbative manner. 
For the definite discussion, we assume that 
the self-force begins to act on the orbit at $\lambda=0$, 
and find that the perturbative evolution is correct 
at $\lambda=O(\mu^\alpha),\,0\geq\alpha>-1/2$. 
Using this result, we first consider to rewrite the orbital equation 
in a numerically convenient form. 

When the orbit is a geodesic, $r$- and $\theta$-motions are periodic. 
Then it is convenient to describe a geodesic 
with phase functions $\chi_r$ and $\chi_\theta$ as 
\begin{eqnarray}
R({\cal E},\chi_r) &=& \sum_n R^{(n)}({\cal E})e^{i2\pi n \chi_r}
\,, \quad 
{d\over d\lambda}\chi_r \,=\, {1\over\lambda_r({\cal E})} 
\,, \\ 
\Theta({\cal E},\chi_\theta) &=& 
\sum_n \Theta^{(n)}({\cal E})e^{i2\pi n \chi_\theta} 
\,, \quad 
{d\over d\lambda}\chi_\theta \,=\, {1\over\lambda_\theta({\cal E})} 
\,.
\end{eqnarray}
We consider the similar type of description when we consider the self-force. 
One can rewrite (\ref{eq:d_r}) and (\ref{eq:d_the}) as 
\begin{eqnarray} 
\delta r &=& \bar R_i\delta {\cal E}^i+\bar R_\chi \delta\chi_r
\,, \quad 
\bar R_i \,=\, \sum_n R^{(n)}_{,i}e^{i2\pi n \chi_{r0}}
\,, \quad 
\bar R_\chi \,=\, \sum_n i2\pi nR^{(n)}e^{i2\pi n \chi_{r0}} 
\,, \label{eq:d_r_x} \\ 
\delta\theta 
&=& \bar\Theta_i\delta {\cal E}^i+\bar\Theta_\chi \delta\chi_\theta 
\,, \quad 
\bar\Theta_i \,=\, \sum_n \Theta^{(n)}_{,i}e^{i2\pi n \chi_{\theta 0}} 
\,, \quad 
\bar\Theta_\chi \,=\, \sum_n i2\pi n\Theta^{(n)}e^{i2\pi n \chi_{\theta 0}} 
\,, \label{eq:d_the_x} 
\end{eqnarray}
where $\chi_{r 0}$ and $\chi_{\theta 0}$ describe 
the phase functions of the initial geodesic, 
and $\delta\chi_r$ and $\delta\chi_\theta$ are 
the deviation by the self-force satisfying 
\begin{eqnarray}
\delta\chi_r &=& 
-{\lambda-\bar\lambda_{r0}\over\lambda_r}
{\lambda_{r,i}\over\lambda_r}\delta{\cal E}^i
-{\delta\bar\lambda_{r0}\over\lambda_r}
\,, \quad 
\delta\chi_\theta \,=\, 
-{\lambda-\bar\lambda_{\theta 0}\over\lambda_\theta}
{\lambda_{\theta,i}\over\lambda_\theta}\delta{\cal E}^i
-{\delta\bar\lambda_{\theta 0}\over\lambda_\theta}
\,. \label{eq:chi_r_the} 
\end{eqnarray}
We find that, with the effect of the self-force, 
the phase functions satisfy 
\begin{eqnarray}
{d\over d\lambda}(\chi_r+\delta\chi_{r0}) &=& 
{1\over\lambda_r+\lambda_{r,i}\delta{\cal E}^i} 
+F_{\chi_r}({\cal E}_0,\chi_{r 0},\chi_{\theta 0}) 
\,, \quad 
F_{\chi_r} \,=\, -{1\over \lambda_r}\sum_{m,n}R_e^{(m,n)}
e^{i2\pi(m\chi_{r 0}+n\chi_{\theta 0})} 
\,, \\ 
{d\over d\lambda}(\chi_\theta+\delta\chi_{\theta 0}) &=& 
{1\over\lambda_\theta+\lambda_{\theta,i}\delta{\cal E}^i} 
+F_{\chi_\theta}({\cal E}_0,\chi_{r 0},\chi_{\theta 0}) 
\,, \quad 
F_{\chi_\theta} \,=\, -{1\over \lambda_r}\sum_{m,n}\Theta_e^{(m,n)}
e^{i2\pi(m\chi_{r 0}+n\chi_{\theta 0})} 
\,.
\end{eqnarray}

One can see that (\ref{eq:chi_r_the}) becomes that of a geodesic 
when we switch off the self-force, 
thus, by the renormalized perturbation method, 
we formally obtain the orbital evolution eqution with the self-force as 
\begin{eqnarray}
{d\over d\lambda}{\cal E} &=& F_{\cal E}({\cal E},\chi_r,\chi_\theta) 
\,, \quad 
{d\over d\lambda}\chi_r \,=\, {1\over\lambda_r({\cal E})}
+F_{\chi_r}({\cal E},\chi_r,\chi_\theta) 
\,, \quad 
{d\over d\lambda}\chi_\theta \,=\, {1\over\lambda_\theta({\cal E})}
+F_{\chi_\theta}({\cal E},\chi_r,\chi_\theta) 
\,, \label{eq:full1} \\ 
r &=& \sum_n R^{(n)}({\cal E})e^{i2\pi n \chi_r}
\,, \quad 
\theta \,=\, \sum_n \Theta^{(n)}({\cal E})e^{i2\pi n \chi_\theta}
\,, \quad 
{d\over d\lambda}t \,=\, X({\cal E},r,\theta) 
\,, \quad 
{d\over d\lambda}\phi \,=\, Y({\cal E},r,\theta) 
\,, \label{eq:full2} 
\end{eqnarray}
where $F_{\cal E}$ is given 
by (\ref{eq:g_E_f}), (\ref{eq:g_L_f}) and (\ref{eq:g_K_f}). 

The evaluation of (\ref{eq:full1}) and (\ref{eq:full2}) may be possible 
by a future investigation of a regularization calculation\cite{reg}. 
But, the calculation may be complex and costly 
unless we have a new breakthrough 
in the regularization calculation strategy. 
We, therefore, introduce an approximate calculation 
using only (\ref{eq:galtsov}), of which 
we already have a well-established calculation technique 
and a number of results\cite{gw}. 

From the perturbative result 
in Subsec.\ref{sec:r-the} and Subsec.\ref{sec:t-p}, 
the orbital evolution is dominantely described as 
\begin{eqnarray}
\delta{\cal E} &=& \lambda \dot{\cal E}^{R-ret.(0,0)\,}({\cal E}_0)
\,, \quad 
\delta\bar\lambda_r \,=\, -{\lambda^2\over 2}
{\lambda_{r,i}\over\lambda_r}\dot{\cal E}^{R-ret.\,(0,0)\,i}({\cal E}_0)
\,, \quad 
\delta\bar\lambda_\theta \,=\, -{\lambda^2\over 2}
{\lambda_{\theta,i}\over\lambda_\theta}
\dot{\cal E}^{R-ret.\,(0,0)\,i}({\cal E}_0)
\,, \\ 
\delta \bar t &=& -{\lambda^2\over 2}
\dot T_{,i}\dot{\cal E}^{R-ret.\,(0,0)\,i}({\cal E}_0) 
\,, \quad 
\delta \bar \phi \,=\, -{\lambda^2\over 2}
\dot\Phi_{,i}\dot{\cal E}^{R-ret.\,(0,0)\,i}({\cal E}_0) 
\,. 
\end{eqnarray}
We call the calculation using these dominant parts 
by an approximate adiabatic calculation, 
and the evolution equation becomes 
\begin{eqnarray}
{d\over d\lambda}{\cal E}^{adi.} &=& \dot{\cal E}^{R-ret.}({\cal E}^{adi.}) 
\,, \quad 
{d\over d\lambda}\chi_r^{adi.} \,=\, {1\over\lambda_r({\cal E}^{adi.})}
\,, \quad 
{d\over d\lambda}\chi_\theta^{adi.} 
\,=\, {1\over\lambda_\theta({\cal E}^{adi.})}
\,, \label{eq:full1_x} \\ 
r^{adi.} &=& \sum_n R^{(n)}({\cal E}^{adi.})e^{i2\pi n \chi_r^{adi.}}
\,, \quad 
\theta^{adi.} \,=\, \sum_n \Theta^{(n)}({\cal E}^{adi.})
e^{i2\pi n \chi_\theta^{adi.}}
\,, \\ 
{d\over d\lambda}t^{adi.} &=& X({\cal E}^{adi.},r^{adi.},\theta^{adi.}) 
\,, \quad 
{d\over d\lambda}\phi^{adi.} \,=\, Y({\cal E}^{adi.},r^{adi.},\theta^{adi.}) 
\,. \label{eq:full2_x} 
\end{eqnarray}

For a practical use of this approximate adiabatic calculation, 
we make a rough estimate on how correctly 
the orbit can be predicted by this method. 
We suppose to divide the whole domain of $\lambda$ 
into an infinite number of domains $\lambda_k<\lambda<\lambda_{k+1}$ 
whose interval is $O(\mu^\alpha),\,\alpha\to -1/2$. 
Then we may apply 
our perturbative analysis of the orbital evolution at each domain. 
We define the orbital `constants' at $\lambda=\lambda_k$ 
as $\{{\cal E}^{(k)},\,\bar\lambda_r^{(k)},\,
\bar\lambda_\theta^{(k)},\,\bar t^{(k)},\,\bar\phi^{(k)}\}$. 
Again, for a definite discussion, we assume that 
the self-force begins to act at $\lambda=\lambda_0$. 

By the result of Subsec.\ref{sec:r-the} and Subsec.\ref{sec:t-p}, 
a dominant contribution makes the evolution of these `constants' as 
\begin{eqnarray}
{\cal E}^{(k+1)}-{\cal E}^{(k)} &\sim& 
O(\mu^{1+\alpha}) \,, \quad 
\bar\lambda_r^{(k+1)}-\bar\lambda_r^{(k)} \,\sim\, 
O(\mu^{1+2\alpha}) \,, \quad 
\bar\lambda_\theta^{(k+1)}-\bar\lambda_\theta^{(k)} \,\sim\, 
O(\mu^{1+2\alpha}) \,, \nonumber \\ 
\bar t^{(k+1)}-\bar t^{(k)} &\sim& 
O(\mu^{1+2\alpha}) \,, \quad 
\bar\phi^{(k+1)}-\bar\phi^{(k)} \,\sim\, 
O(\mu^{1+2\alpha}) \,. 
\end{eqnarray}
After passing by $N\sim O(\mu^\beta)$ finite domains, we have 
\begin{eqnarray}
{\cal E}^{(N)} &\sim& O(\mu^{1+\alpha+\beta})
\,, \quad 
\bar\lambda_r^{(N)} \,\sim\, O(\mu^{1+2\alpha+\beta}) 
\,, \quad 
\bar\lambda_\theta^{(N)} \,\sim\, O(\mu^{1+2\alpha+\beta}) 
\,, \nonumber \\ 
\bar t^{(N)} &\sim& O(\mu^{1+2\alpha+\beta})
\,, \quad 
\bar\phi^{(N)} \,\sim\, O(\mu^{1+2\alpha+\beta}) 
\,. 
\end{eqnarray}

On the other hand, 
the part we ignore for the approximate adiabatic calculation 
contaminates the evolution as 
\begin{eqnarray}
\delta({\cal E}^{(k+1)}-{\cal E}^{(k)}) &\sim& O(\mu)
\,, \quad 
\delta(\bar\lambda_r^{(k+1)}-\bar\lambda_r^{(k)}) \,\sim\, O(\mu^{1+\alpha}) 
\,, \quad 
\delta(\bar\lambda_\theta^{(k+1)}-\bar\lambda_\theta^{(k)}) 
\,\sim\, O(\mu^{1+\alpha}) 
\,, \nonumber \\ 
\delta(\bar t^{(k+1)}-\bar t^{(k)}) &\sim& O(\mu^{1+\alpha})
\,, \quad 
\delta(\bar\phi^{(k+1)}-\bar\phi^{(k)}) \,\sim\, O(\mu^{1+\alpha}) 
\,. 
\end{eqnarray}
We assume that the ignored part affects the evolution 
like an randam Gaussian noise 
and estmate the error as 
\begin{eqnarray}
\delta{\cal E}^{(N)} &\sim& O(\mu^{1+\beta/2})
\,, \quad 
\delta\bar\lambda_r^{(N)} \,\sim\, O(\mu^{1+\alpha+\beta/2}) 
\,, \quad 
\delta\bar\lambda_\theta^{(N)} \,\sim\, O(\mu^{1+\alpha+\beta/2}) 
\,, \nonumber \\ 
\delta\bar t^{(N)} &\sim& O(\mu^{1+\alpha+\beta/2})
\,, \quad 
\delta\bar\phi^{(N)} \,\sim\, O(\mu^{1+\alpha+\beta/2}) 
\,. 
\end{eqnarray}

Using this estimation, 
we discuss the predictability of the approximate adiabatic calculate. 
The error of the rotation per time $\phi/t$ is estimated as 
\begin{eqnarray}
\delta\left({\phi\over t}\right) \sim O(\mu^{-\alpha-\beta/2}) \,. 
\end{eqnarray}
For an accurate prediction, we require $\delta(\phi/t)<1$ 
and we have $\beta<-2\alpha \to 1$, 
which is always satisfied since $N$ is just an integer. 

The error of the rotation phase $\phi$ is estimated as 
\begin{eqnarray}
\delta\phi \sim O(\mu^{1+\alpha+\beta/2}) \,. 
\end{eqnarray}
For a correct prediction of the phase, we have $\beta>-2\alpha-2 \to -1$. 
This shows that we have a prediction of the rotation phase 
only at $\lambda \leq O(\mu^{-3/2})$.

\subsection{Gauge Issue} \label{sec:gauge}

The gravitational self-force problem has an exceptional difficulty 
because of the so-called gauge problem\cite{gauge}. 
The regularization formulation was originally formulated 
in the harmonic gauge condition\cite{g_self,g_reg}, 
and we have the divergent S-part only in the harmonic gauge. 
It was pointed out that, 
when we subtract the S-part 
from the full metric perturbation in the radiation gauge\cite{chrz}, 
we have a divergent residue\cite{gauge} 
because of the divergent gauge transformation at the particle location. 
For this reason, 
the calculation of the full metric perturbation 
in the harmonic gauge condition, 
or the calculation of the S-part of the metric perturbation 
in the radiation gauge 
become important issues in calculating 
the gravitational self-force in a Kerr background. 

On the other hand, in the approximate adiabatic calculation, 
all we need is to evaluate (\ref{eq:galtsov}), 
which is proven to agree 
with the radiation reaction calculation\cite{galtsov} in part. 
A number of previous works\cite{gw} prove 
there is no divergence in the approximate adiabatic calculation. 
The investigation in Subsec.\ref{sec:react} shows that 
the self-force derived by the radiative Green function 
corresponds to the two-point averaged self-force 
derived by the R-part of the retarded Green function. 
We consider that, by taking a two point average of the self-force, 
the divergent S-part vanishes as in (\ref{eq:rad}). 
We consider that, in the approximate adiabatic calculation, 
the same cancellation mechanism happens 
including divergence by the gauge transformation. 
But there still be an abiguity of a finite gauge choice. 
Here we prove the result by the approximate adiabatic calculation 
is actually gauge-invariant 
by showing the gauge invariance of (\ref{eq:galtsov}). 

Using a Killing vector $\eta_\alpha$ 
and a Killing-Yano tensor $\eta_{\alpha\beta}$ of a Kerr spacetime, 
conserved quantities along a geodesic are described as 
\begin{eqnarray}
E &=& \eta_\alpha v^\alpha \,, \quad 
Q \,=\, \eta_{\alpha\beta} v^\alpha v^\beta \,, 
\end{eqnarray}
where $v^\alpha(\tau)=dz^\alpha/d\tau$ is 4-velocity 
and $\tau$ is proper time of the geodesic. 
(\ref{eq:galtsov}) can be rewritten as 
\begin{eqnarray}
\lim_{T\to\infty}{1\over 2T}\int^T_{-T}d\tau{d\over d\tau}E 
&=& \lim_{T\to\infty}{1\over 2T}\int^T_{-T}d\tau
\eta_\alpha(z(\tau)) F^\alpha(\tau) 
\,, \label{eq:react_e} \\ 
\lim_{T\to\infty}{1\over 2T}\int^T_{-T}d\tau{d\over d\tau}Q 
&=& \lim_{T\to\infty}{1\over 2T}\int^T_{-T}d\tau
2\eta_{\alpha\beta}(z(\tau)) v^\alpha(\tau) F^\beta(\tau) 
\,, \label{eq:react_q} 
\end{eqnarray}
where $F^\beta$ is the self-force vector. 

By a gauge transformation $x^\alpha\to x^\alpha+\xi^\alpha$, 
the self-force is transformed as 
\begin{eqnarray}
F^\alpha(\tau) &\rightarrow& F^\alpha(\tau)+\delta_\xi F^\alpha(\tau) 
\,, \\ 
\delta_\xi F^\alpha(\tau) &=& 
-\Bigl(v^\beta(\tau) v^\gamma(\tau) \xi^\alpha{}_{;\beta\gamma}(z(\tau))
+R^\alpha{}_{\beta\gamma\delta}(z(\tau))
v^\beta(\tau) \xi^\gamma(z(\tau)) v^\delta(\tau) \Bigr) 
\,. 
\end{eqnarray}
The increment of (\ref{eq:react_e}) and (\ref{eq:react_q}) 
by this extra term $\delta_\xi F^\alpha(\tau)$ becomes 
\begin{eqnarray}
\lim_{T\to\infty}{1\over 2T}\int^T_{-T}d\tau
\delta\left[{d\over d\tau}E(\tau)\right] 
&=& \lim_{T\to\infty}{1\over 2T}
\biggl[-\eta^\alpha v^\beta \xi_{\alpha;\beta}
+v^\beta \eta^\alpha{}_{;\beta} \xi_\alpha\biggr]^T_{-T}
\,, \label{eq:E-gauge} \\ 
\lim_{T\to\infty}{1\over 2T}\int^T_{-T}d\tau
\delta\left[{d\over d\tau}Q(\tau)\right] 
&=& \lim_{T\to\infty}{1\over 2T}
\biggl[-2v_\beta \eta^{\alpha\beta} v^\gamma \xi_{\alpha;\gamma}
+2v_\beta v^\gamma \eta^{\alpha\beta}{}_{;\gamma} \xi_\alpha\biggr]^T_{-T}
\,, \label{eq:Q-gauge} 
\end{eqnarray}
where we use Killing equations 
$\eta_{\alpha;\beta\gamma}=\eta_\delta R^\delta{}_{\gamma\beta\alpha}$ 
and $\eta_{\alpha\beta;\gamma\delta}
=\eta_{\epsilon\beta} R^\epsilon{}_{\delta\gamma\alpha} 
+\eta_{\alpha\epsilon} R^\epsilon{}_{\delta\gamma\beta}$. 
Since the gauge dependence is totally integrated out, 
the gauge dependence of the approximate adiabatic calculation 
vanishes by taking $T\to \infty$. 

We comment that 
the gauge dependence of (\ref{eq:full1}) and (\ref{eq:full2}) 
is highly non-trivial 
and we left it as a future problem.

\section{Summary}

In this paper, 
we discuss a method to calculate an orbital evolution 
by a scalar/electromagnetic/gravitational self-force. 
We assume that the self-force is weak 
and that the orbit can be approximated by a geodesic 
at each instant of time, 
with which one can derive the self-force. 
We note that the geodesic equation is 
a set of second-order differential equations of four components, 
and we have 7 integral constants, 
$E$, $L$, $K$, $\lambda_0$, $\lambda_1$, $t_0$ and $\phi_0$. 
Instead of calculating the orbit itself, 
we derive the equations of these `constants' by the self-force 
under this assumption. 

We first consider the evolution of $E$, $L$ and $K$ 
since the evolution equations of these `constants' are 
directly derived by the self-force vector. 
We exploit the symmetry of a Kerr spacetime 
together with a family of geodesics, which induce the self-force. 
By applying the symmetry transformation to the self-force vector, 
we find that the time-averaged evolution of $E$, $L$ and $K$ 
can be derived by using a radiative Green function, 
which has a number of technical advantage in practice. 
However, we also find that the orbit does not evolve 
adiabatically in an exact sense. 

In order to understand the orbital evolution by the self-force, 
we next consider the orbital equation in a perturbative manner. 
We integrate the orbital equation by a time scale, 
sufficiently long, but, less than the dephasing time 
when the linear perturbation of the orbit becomes invalid. 
Since the orbit does not evolve in an adiabatic manner, 
the orbital constants of a geodesic 
$\{{\cal E},\bar\lambda_r,\bar\lambda_\theta,\bar t,\bar\phi_0\}$ 
are oscillating by the self-force. 
However, we could find out secularly growing parts 
which will dominate the orbital evolution. 
By taking these growing parts only, 
we define an approximate orbital eequation, 
which we call an approximate adiabatic calculation. 
We consider that 
the approximate adiabatic calculation is enough implementable 
by a well-established method 
since it only uses the radiative Green function. 

We also discuss how approximate an orbital evolution can be obtained 
by this calculation method. 
We find that, during the time $O(\mu^{-3/2})$, 
it gives an accurate rotation phase. 
For example, when $10$ solar-mass black hole is inspiralling 
into $10^7$ solar-mass supermassive black hole, 
this corresponds to around $10^9$-rotation period. 
Though the accuracy in predicting a wave form is not clear in our estimate, 
if only the accurate prediction of the rotation phase is important 
for the future LISA observation, 
the approximation proposed here may give a sufficient imformation 
in this case. 

Finally we prove that 
the approximate adiabatic calculation gives a gauge invariant prediction, 
thus, the result is consistent with that by another possible method 
within the approximation scheme.


\noindent

\section*{Acknowledgements}

We thank Professor Clifford Will for encouragement. 
Special thanks to Professor Misao Sasaki and Professor Takahiro Tanaka 
for encouragement and a fruitful discussion with a hospitality 
during the stay in Osaka University and Kyoto University. 
We also thank Dr. Jhingan Sanjay and Dr. Hiroyuki Nakano for a discussion. 
This work was supported in part by the National Science Foundation 
under grant number PHY 00-96522. 


\begin{appendix}

\section*{Circular or Equatorial orbit}

When the orbit is circular, or equatorial, 
we have $dr_0/d\lambda=V({\cal E}_0,r_0)=V({\cal E}_0,r_0)_{,r}=0$, 
or $d\theta_0/d\lambda=U({\cal E}_0,\theta_0)
=U({\cal E}_0,\theta_0)_{,\theta}=0$. 
Then (\ref{eq:d_r_the0}) becomes trivial 
and we cannot evaluate the perturbed orbital equation. 
Here we consider the orbital evolution in these special cases 
by taking the circular, or equatorial limit of a general orbit. 
Before the discussion, 
we note that, by repeating the symmetry argument in Sec.\ref{sec:sym}, 
the self-force acting on $E$, $L$ and $K$ becomes 
\begin{eqnarray}
\left[{d\over d\lambda}{\cal E}\right]^{R-ret./R-adv./rad.}
&=& \displaystyle\cases{
\sum_n \dot {\cal E}^{R-ret./R-adv./rad.(0,n)}(E,L,K) 
e^{i2\pi n{\lambda-\bar\lambda_\theta\over\lambda_\theta}} \,, & 
circular orbit \,, \cr 
\sum_m \dot {\cal E}^{R-ret./R-adv./rad.(m,0)}(E,L,K) 
e^{i2\pi m{\lambda-\bar\lambda_r\over\lambda_r}} \,, & 
equatorial orbit \,, \cr 
\dot {\cal E}^{R-ret./R-adv./rad.(0,0)}(E,L,K) \,, & 
circular and equatorial orbit \,. }
\end{eqnarray}

We consider $r$-motion 
when we consider a self-force acting on a circular orbit. 
The perturbation equation (\ref{eq:d_r_the0}) becomes trivial 
since we have $V_{,i}=V_{,r}=0$ along the orbit by using (\ref{eq:vv}). 
Instead of calculating the orbital equation, 
we consider to take the circular limit of (\ref{eq:d_r}). 
The first term of (\ref{eq:d_r}) is given in (\ref{eq:d_r_the1}) 
and behaves in a regular manner when we take a circular limit. 
On the other hand, 
the second term of (\ref{eq:d_r}) is given in (\ref{eq:d_r_the2}). 
By the regularization calculation starting from (\ref{eq:d_r3}), 
the integration is still finite in the circular limit, 
and $dr_0/d\lambda$ vanishes. 
Thus, we have 
\begin{eqnarray} 
\delta r &=& R_{,i}\delta {\cal E}^i \,, 
\end{eqnarray}
which means the circular orbit stays circular under the self-force 
and the orbit is solely determined by $E$, $L$ and $K$
\footnote{The same proof was given in Ref.\cite{circular}.} 
As a result, the orbital equation becomes 
\begin{eqnarray}
{d\over d\lambda}{\cal E} &=& F_{\cal E}({\cal E},\chi_\theta) 
\,, \quad 
{d\over d\lambda}\chi_\theta \,=\, {1\over\lambda_\theta({\cal E})}
+F_{\chi_\theta}({\cal E},\chi_r,\chi_\theta) 
\,, \\ 
r &=& R^{(0)}({\cal E}) 
\,, \quad 
\theta \,=\, \sum_n \Theta^{(n)}({\cal E})e^{i2\pi n \chi_\theta}
\,, \quad 
{d\over d\lambda}t \,=\, X({\cal E},r,\theta) 
\,, \quad 
{d\over d\lambda}\phi \,=\, Y({\cal E},r,\theta) 
\,, 
\end{eqnarray}
and, under the approximate adiabatic calculation, we have 
\begin{eqnarray}
{d\over d\lambda}{\cal E}^{adi.} &=& \dot{\cal E}^{R-ret.}({\cal E}^{adi.}) 
\,, \quad 
{d\over d\lambda}\chi_\theta^{adi.} 
\,=\, {1\over\lambda_\theta({\cal E}^{adi.})}
\,, \\ 
r^{adi.} &=& R^{(0)}({\cal E}^{adi.})
\,, \quad 
\theta^{adi.} \,=\, \sum_n \Theta^{(n)}({\cal E}^{adi.})
e^{i2\pi n \chi_\theta^{adi.}}
\,, \\ 
{d\over d\lambda}t^{adi.} &=& X({\cal E}^{adi.},r^{adi.},\theta^{adi.}) 
\,, \quad 
{d\over d\lambda}\phi^{adi.} \,=\, Y({\cal E}^{adi.},r^{adi.},\theta^{adi.}) 
\,. 
\end{eqnarray}

Similarly, $\theta$-motion can be drived such that 
the equatorial motion stays equatorial. 
One can also prove by using the symmetry $\theta\to -\theta$. 
Using this symmetry property, when the orbit is equatorial, 
the self-force satisfies $F_\theta=0$, 
and we have $\theta=\pi/2,\,d\theta/d\lambda=0$. 
The orbital equation becomes 
\begin{eqnarray}
{d\over d\lambda}{\cal E} &=& F_{\cal E}({\cal E},\chi_r) 
\,, \quad 
{d\over d\lambda}\chi_r \,=\, {1\over\lambda_\theta({\cal E})}
+F_{\chi_\theta}({\cal E},\chi_\theta) 
\,, \\ 
r &=& \sum_n R^{(n)}({\cal E})e^{i2\pi n \chi_r} 
\,, \quad 
\theta \,=\, {\pi\over 2}
\,, \quad 
{d\over d\lambda}t \,=\, X({\cal E},r,\pi/2) 
\,, \quad 
{d\over d\lambda}\phi \,=\, Y({\cal E},r,\pi/2) 
\,, 
\end{eqnarray}
and, under the approximate adiabatic calculation, we have 
\begin{eqnarray}
{d\over d\lambda}{\cal E}^{adi.} &=& \dot{\cal E}^{R-ret.}({\cal E}^{adi.}) 
\,, \quad 
{d\over d\lambda}\chi_r^{adi.} 
\,=\, {1\over\lambda_r({\cal E}^{adi.})}
\,, \\ 
r^{adi.} &=& \sum_n R^{(n)}({\cal E}^{adi.})e^{i2\pi n \chi_r^{adi.}}
\,, \quad 
\theta^{adi.} \,=\, \pi/2
\,, \\ 
{d\over d\lambda}t^{adi.} &=& X({\cal E}^{adi.},r^{adi.},\pi/2) 
\,, \quad 
{d\over d\lambda}\phi^{adi.} \,=\, Y({\cal E}^{adi.},r^{adi.},\pi/2) 
\,. 
\end{eqnarray}

When the orbit is circular and equatorial, 
the oribtal equation becomes 
\begin{eqnarray}
{d\over d\lambda}{\cal E} &=& F_{\cal E}({\cal E}) 
\,, \\  
r &=& R^{(0)}({\cal E})
\,, \quad 
\theta \,=\, {\pi\over 2}
\,, \quad 
{d\over d\lambda}t \,=\, X({\cal E},r,\pi/2) 
\,, \quad 
{d\over d\lambda}\phi \,=\, Y({\cal E},r,\pi/2) 
\,, 
\end{eqnarray}
and the orbit evolves adiabatically in an exact sense. 

\end{appendix}



\end{document}